\documentclass[12pt]{iopart}
\usepackage{epsfig}
\usepackage{amscd,amssymb}
\usepackage{graphicx}
\usepackage{color}

\begin{document}

\newcommand \be  {\begin{equation}}
\newcommand \bea {\begin{eqnarray} \nonumber }
\newcommand \ee  {\end{equation}}
\newcommand \eea {\end{eqnarray}}

\title[Freezing Transitions and Extreme Values]
{Freezing Transitions and Extreme Values: Random Matrix Theory, $\zeta(1/2+it)$, and Disordered Landscapes}

\vskip 0.2cm
\author{Yan V. Fyodorov$^1$ and Jonathan P. Keating$^2$}
\noindent\small{$^1$ Queen Mary University of London, School of Mathematical Sciences, London E1 4NS, UK}\\
\noindent\small{$^2$ School of Mathematics,
University of Bristol, Bristol BS8 1TW, UK}

\maketitle

\begin{abstract}
We argue that the {\em freezing transition scenario}, previously
%( {\sf Fyodorov $\&$ Bouchaud 2008; Fyodorov, Le Doussal, $\&$ Rosso 2009})
conjectured to occur in the statistical mechanics of $1/f-$noise random energy models,
governs, after reinterpretation, the value distribution of the maximum of the modulus of the characteristic polynomials $p_N(\theta)$ of large
$N\times N$ random unitary (CUE) matrices $U_N$; i.e.~the extreme value statistics of $p_N(\theta)$ when $N \rightarrow\infty$.  In addition, we argue that it leads to multifractal-like behaviour in the total length $\mu_N(x)$ of the intervals in which $|p_N(\theta)|>N^x, x>0$, in the same limit.  We speculate that our results extend to the large values taken by the Riemann zeta-function $\zeta(s)$ over stretches of the critical line $s=1/2+it$ of given constant length, and present the results of numerical computations of the large values of $\zeta(1/2+it)$.  Our main purpose is to draw attention to the unexpected connections between these different extreme value problems.

\end{abstract}

\section{Introduction}

Over the past 40 years, considerable evidence has accumulated for connections between certain statistical properties of the Riemann zeta-function, $\zeta(s)$, and those of large random matrices.   For example, correlations between the nontrivial zeros of the zeta function on the {\em critical line} $s=1/2+it, \, t\in \mathbb{R},$ are believed to coincide, in the limit $t \rightarrow \infty$, with those between the eigenvalues of large random unitary or hermitian matrices, and the value distribution of $\zeta(1/2+it)$ is believed to be related to that of the characteristic polynomials of large random unitary or hermitian matrices.  Our purpose here is to connect these two areas of research to a third, the statistical mechanics of disordered energy landscapes.  The analogy we develop suggests that the freezing transition observed in the statistical mechanical problem also governs the extreme values taken by the characteristic polynomials of random matrices and the zeta function.  This sheds new light on the longstanding problem of determining the maximum size of the zeta function.

The Riemann zeta-function
\be\label{zetadef}
\zeta(s)=\sum_{n=1}^{\infty}\frac{1}{n^s}=\prod_{p}\left(1-\frac{1}{p^s}\right)^{-1}
\ee
is of central importance in mathematics because it encodes the distribution of the primes $p$ in the positions of its {\em non-trivial zeros}.  The Riemann Hypothesis places these zeros on the critical line.  Some of the most important questions in the theory of the zeta function concern the distribution of values it takes on the critical line.  It was proved by Selberg \cite{Sel44, Sel46}, for example, that, when $t\in \mathbb{R}$, $\log |\zeta(1/2+it)|$
satisfies a central limit theorem:
\be
\label{CLT}
\fl \lim_{T\rightarrow\infty}\frac{1}{T}{\rm meas}\left\{T\le t\le 2T:\alpha\le\frac{\log |\zeta(1/2+it)|}{\sqrt{\frac{1}{2}\log\log\frac{t}{2\pi}}}\le\beta\right\}
=\frac{1}{\sqrt{2\pi}}\int_{\alpha}^{\beta}\exp(-x^2/2)dx
\ee
which implies that its typical size is of the order of $\sqrt{\log\log t}$ when $t \rightarrow \infty$ (see, e.g.,  \cite{Titch, Laurin}).  As regards the exceptionally large values taken by the zeta function over long ranges, the Lindel{\" o}f hypothesis asserts that $|\zeta(1/2+it)|=o(t^\epsilon)$ for any $\epsilon>0$, the Riemann Hypothesis implies that
\be\label{Ozeta}
|\zeta(1/2+it)|=O\left(\exp\left(\frac{c_1\log t}{\log\log t}\right)\right),
\ee
where $c_1$ is a constant, and, unconditionally, we know that
\be\label{Omegazeta}
|\zeta(1/2+it)|=\Omega\left(\exp\left(\sqrt{\frac{\log t}{\log\log t}}\right)\right),
\ee
meaning that $|\zeta(1/2+it)|$ takes the value in the argument on the right-hand side infinitely often (see, e.g., \cite{Titch}).  The exceptionally large values of $|\zeta(1/2+it)|$ thus lie in the range between (\ref{Ozeta}) and (\ref{Omegazeta}).  The problem of determining where precisely within this range they lie has attracted considerable attention in recent years, but remains unresolved.  The extreme values in question are so rare that extensive numerical computations have thus far failed to settle the matter.

It was observed by Montgomery (in relation to a conjecture, to be described below, of Farmer, Gonek $\&$ Hughes, see \cite{FGH}) that treating the local maxima of $\log|\zeta(1/2+it)|$ as being statistically independent, assuming that their values satisfy the central limit theorem (\ref{CLT}), and using the fact the the number of local maxima of $|\zeta(1/2+it)|$ in $0\le t\le T$ is of the order of $\frac{T}{2\pi}\log\frac{T}{2\pi}$ (the number of zeros in the range), implies that the typical size of the maximum value of $|\zeta(1/2+it)|$ is of the order of $\exp\left(c_2\sqrt{\log (t) \log\log (t)}\right)$, where $c_2$ is a constant\footnote{As will be discussed in Section 3, the typical size of the maximum of $M$ i.i.d.~normal,  mean-zero random variables $V_i$  with variance $\mathbb{E}\{V_i^2\}=\sigma<\infty$ behaves asymptotically like  $a_M\approx \sqrt{2\sigma \log{M}}$. The estimate for the maximum modulus of the $\zeta-$function follows from considering $M=\frac{t}{2\pi}\log\frac{t}{2\pi}$ samples drawn independently from the gaussian distribution in (\ref{CLT}) with variance $\sigma=\frac{1}{2} \log\log{(t/2\pi)}$.}.  Note that this is considerably larger than the typical size of $|\zeta(1/2+it)|$, and that it is closer to (\ref{Omegazeta}) than to (\ref{Ozeta}), implying that the extreme values are not much larger than the largest value known to be reached infinitely often. Significantly, this calculation makes clear that the square root in the estimate of the maximum size is related to the fact that the exponential in the integrand on the right hand side of (\ref{CLT}) is quadratic in $x$.  Had the exponential been linear in $x$ then the result would have been closer to the upper limit  (\ref{Ozeta}).  It is worth remarking that the Montgomery model would predict a Gumbel distribution (see Section 3) for the fluctuations of the extreme values of $|\zeta(1/2+it)|$ around this typical size.

It is important to note that Montgomery's observation rests on two key assumptions.  The first is that the central limit theorem (\ref{CLT}) extends out to the range of the large values predicted.  This is far beyond the range for which it has been established.  The second is that the values of the local maxima of $|\zeta(1/2+it)|$ are uncorrelated.   In fact, the values of $\log|\zeta(1/2+it)|$ are known to be correlated in a way that is significant from the point of view of the ideas we shall  explore here.  Specifically,  let us define, for a fixed $t\in \mathbb{R}$,
\be\label{logzetaint}
 V_t^{(\zeta)}(x)=-2\log{|\zeta\left(\frac{1}{2}+i(t+x)\right)|}=-2\mbox{Re}\log{\zeta\left(\frac{1}{2}+i(t+x)\right)}.
\ee
(The factor of $-2$ is introduced for reasons to be explained below.)
The central limit theorem (\ref{CLT}) implies that, when $t\to \infty$, $V_t^{(\zeta)}(x)$
behaves like a Gaussian random function of  $x$. To characterize such a random process it is natural to consider the two-point correlation function $\left\langle V^{(\zeta)}_t(x_1)V^{(\zeta)}_t(x_2) \right\rangle$, with brackets
$\left\langle ... \right\rangle$ denoting the average over an interval $[t-h/2,t+h/2]$ such that $\frac{1}{\log{t}}\ll h \ll t $.
A simple argument, sketched in Appendix A, shows that when $t\rightarrow\infty$
\be\label{corzetadiag3int}
%  \left\langle  V_t^{(\zeta)}(x_1)V_t^{(\zeta)}(x_2) \right\rangle\approx \left\{ \begin{array}{c} -\frac{1}{2}\log{|x_1-x_2|}, \,\, \mbox{for} \quad  \frac{1}{\log{t}}\ll |x_1-x_2|\ll 1\\
\left\langle  V_t^{(\zeta)}(x_1)V_t^{(\zeta)}(x_2) \right\rangle\approx \left\{ \begin{array}{c} -2\log{|x_1-x_2|}, \,\, \mbox{for} \quad  \frac{1}{\log{t}}\ll |x_1-x_2|\ll 1\\
%\frac{1}{2}\log{\log{t}}, \,\, \mbox{for} \quad  |x_1-x_2| \ll \frac{1}{\log{t}} \end{array}\right.
2\log{\log{t}}, \,\, \mbox{for} \quad  |x_1-x_2| \ll \frac{1}{\log{t}} \end{array}\right.
\ee
This illustrates the fact that values of $\log|\zeta(1/2+it)|$ are indeed correlated (see also \cite{Bour}).  The significance of the precise form of the correlations will become apparent when we draw comparisons with corresponding problems in random matrix theory and statistical mechanics.

Over the past decade it has become a well-established paradigm that many statistical properties of the Riemann zeta function along the critical line can be understood by comparing them to analogous properties of the characteristic polynomials of random matrices \cite{KS, HKO, CFKRS05, GHK, CFKRS08}. Let $U_N$ be an $N\times N$ unitary matrix, chosen uniformly at random from the unitary group ${\cal U}(N)$ (i.e.~$U_N$ lies in the Circular Unitary Ensemble, or CUE, of random matrices), and denote its eigenvalues by $\exp{(i\phi_1)},\ldots, \exp{(i\phi_N)}$. Let
\be\label{pdef}
p_N(\theta)=\det{\left(1-U_N\,e^{-i\theta}\right)}=\prod_{n=1}^{N}\left(1-e^{i(\phi_n-\theta)}\right)
\ee
be the corresponding characteristic polynomial.  To this end it is instructive to compare $V_t^{(\zeta)}(x)$ from (\ref{logzetaint}) with  $V^{(U)}_N(\theta)=-2\log{|p_N(\theta)|}$.  (Again, the factor of $-2$ is introduced for reasons to be explained below.)
$V^{(U)}_N(\theta)$ satisfies a central limit theorem that is the analogue of (\ref{CLT}) \cite{KS, BF}.  Specifically, the values of $\log|p_N(\theta)|$ normalized by $\sqrt{\frac{1}{2}\log N}$, have a limiting distribution as $N\rightarrow\infty$ given by the right-hand side of (\ref{CLT}).  Identifying the mean density of the eigenvalues, $N/2\pi$ with the mean density of the Riemann zeros near to height $t$,
$\frac{1}{2\pi}\log\frac{t}{2\pi}$, renders the agreement complete.

Importantly for us here, $V^{(U)}_N$ has the following representation \cite{HKO}:
 \be\label{8int}
V^{(U)}_N(\theta)=\sum_{n=1}^{\infty}\frac{1}{\sqrt{n}}\left[e^{-i n\theta}v^{(N)}_n+\mbox{comp. conj.}\right], \quad v^{(N)}_n=\frac{1}{\sqrt{n}}\mbox{Tr}\left(U_N^{n}\right)
\ee
The coefficients $v^{(N)}_n$ for any fixed finite set of integers $n$ tend, in the limit $N\to \infty$, to i.i.d.~complex gaussian variables with zero mean and variance $\mathbb{E}\{|v_n|^2\}=1$ \cite{DiacShahsh}. A simple calculation
 (cf.~(\ref{logcovarcont}, \ref{1/f}) below)  then shows that
 $\mathbb{E}\left\{V^{(U)}_N(\theta_1)V^{(U)}_N(\theta_2)\right\}$ tends in the limit $N\to \infty$ to
 $-2\log{2|\sin{\frac{1}{2}(\theta_1-\theta_2)|}}$, and so exhibits precisely the same logarithmic behaviour at small distances as we found for the zeta-function.  For large but finite $N$, the logarithmic divergence can be shown to saturate at $|\theta_1-\theta_2|\sim N^{-1}$, so after associating $N\sim \log \frac{t}{2\pi}$ the correspondence between $V^{(U)}_N(\theta)$ and $V_t^{(\zeta)}(x)$ becomes complete.  This is significant from the point of view we seek to develop.

Our goal here is to determine the maximum value of $|p_N(\theta)|$ over some specified interval $0\le\theta\le L$, or, more precisely, the distribution of these maximum values when  $U_N $ ranges over ${\cal U}(N)$.  Our second goal is then to use the random-matrix results to motivate predictions for the extreme values of the Riemann zeta function.   The first steps in this direction were taken by Farmer, Gonek and Hughes \cite{FGH}, who determined the tail of the distribution of the maximum values of $|p_N(\theta)|$ when $L=2\pi$.  They modeled $\zeta(1/2+it)$ in the range $0\le t \le T$ by the characteristic polynomials of a set of approximately $T$ independently chosen random matrices (recall that the number of zeros in the range is of the order of $T\log T$, and the identification $N\sim \log{T}$).  This corresponds to asking for the typical size of the maximum value that $|p_N(\theta)|$ takes when  $0\le\theta\le 2\pi$ and  $U_N $ is sampled independently a large (exponentially in $N$) number of times from within ${\cal U}(N)$.  The resulting conjecture is in accord with the Montgomery heuristic (and the random-matrix approach is sufficiently refined to predict a value for the constant $c_2$).  This is not altogether surprising, because, as discussed above, the central limit theorem for $\log{|p_N(\theta)|}$ corresponds to that for $\log|\zeta(1/2+it)|$, and, moreover, in the random matrix case we know that the gaussian distribution extends to the large deviation regime \cite{HKO}.

The focus here will differ from that of \cite{FGH} in the following ways.  We shall be concerned with the maximum values of the characteristic polynomials of single matrices, rather than with large numbers of matrices, and will obtain the full value distribution of the maxima in the limit as $N\rightarrow \infty$, rather than concentrating on the tail that is relevant when maximizing over many matrices.  This leads to a model for the distribution of maximum values of $|\zeta(1/2+it)|$ over  $T \le t \le T+L$, $L \le 2\pi$,  rather than $0 \le t \le T$ as $T\rightarrow\infty$.  Furthermore, it makes the problem of numerical computation of the distribution in question significantly easier, because one is finding the maximum only of $\sim L\log T$ rather than $\sim T \log T$ numbers (the values of the local maxima).

Our main purpose here is to link the problems of finding the extreme value statistics of the characteristic polynomials of random matrices and of  $|\zeta(1/2+it)|$ to an interesting and important class of problems in statistical mechanics.
The maximum value of  $p_N(\theta)$ over the interval in question can be characterized in terms of the moments
\be\label{1}
{\cal Z}_{N}(\beta;L)=\frac{N}{2\pi}\int_0^{L}|p_N(\theta)|^{2\beta}d\theta\equiv \frac{N}{2\pi}\int_0^{L}e^{-\beta V_N(\theta)}\,d\theta, \quad \beta>0
\ee
where $V_N(\theta)=-2\log{|p_N(\theta)|}$.  Specifically, if ${\cal F}(\beta)=-\beta^{-1}\log{{\cal Z}_N(\beta)}$, then
\be\label{2}
\lim_{\beta\to \infty}{\cal  F}(\beta)=\min_{\theta\in(0,L)}V_N(\theta)=\min_{\theta\in(0,L)}\left[-2\log{|p_N(\theta)|}\right]=-2\max_{\theta\in(0,L)}
 \log{|p_N(\theta)|}.
\ee
The key point is that (\ref{1}) takes the form of a partition function for a system with energy
$V_N(\theta)$ and inverse temperature $\beta$, and ${\cal F}(\beta)$ may then be associated with the corresponding free energy.  Recalling that the values of $V_N(\theta)$ are gaussian distributed and logarithmically correlated, it is natural to draw comparisons with a class of problems that has attracted a good deal of attention recently in the area of disordered systems, namely the statistical mechanics of a single particle equilibrated in a random potential energy described by a gaussian  random processes with logarithmic correlations. For example the statistical mechanics of systems in which the energy is given by a random Fourier series, similar to (\ref{8int}) was addressed in \cite{FB08} and \cite{FLeDR09}.  In the statistical mechanical problem there has been a particular focus on the freezing transition which dominates the low temperature limit and determines the extreme value statistics.  We shall argue that a similar freezing transition determines the extreme value statistics of the characteristic polynomials and hence, conjecturally, of $|\zeta(1/2+it)|$, and that therefore the logarithmic correlations exhibited by $V^{(U)}_N(\theta)$ and $V_t^{(\zeta)}(x)$ play an important role.  Explaining this observation represents our main objective.  The implications are wide-ranging, in that they allow us to predict explicit formulae for the extreme value statistics.  It should be emphasized that we see this as the first steps in exploring the connections, and that we are aware of the speculative nature of many of the predictions we shall make.  We see it as a major challenge to explore these ideas more rigorously.  We put forward these speculations in the hope that they will stimulate new directions of research into the long-standing problems we address.

The structure of this paper is as follows.  We summarize our predictions in the next section, and also briefly discuss
 preliminary numerical evidence for $\zeta(1/2+it)$ and random matrices in support of them.  Some of these numerical results were first outlined in our short communication
 \cite{FHK}, which we refer to below as FHK.
As the statistical mechanical problems we shall compare to play a central role in this story, and as they may be unfamiliar in the context of the connections between random matrix theory and number theory, we give an overview of the key concepts and main results in Section 3.  This review of the literature is necessarily lengthy because the intuitions coming from statistical mechanics are essential to explain our ideas.
In Section 4 we show how to employ the statistical mechanics methodology in the context of the extreme value statistics of the characteristic polynomials of random unitary (CUE) matrices.  Finally, in Section 5 we outline briefly how these calculations may be used to motivate predictions for the extreme value statistics for the Riemann zeta-function.

\section{Summary of predictions}

As explained in the introduction, we see our main goal in this paper as being to explain the analogies between the statistical mechanics of certain disordered systems, in particular the freezing transition observed there, and the extreme value statistics of the characteristic polynomials of large random unitary matrices and the Riemann zeta-function.  These analogies lead to a range of predictions, a representative selection of which we list here.  These predictions follow from the calculations outlined in Sections 4 and 5, and are motivated by similar calculations in Statistical Mechanics reviewed in Section 3.

\subsection{Statistics of the moments of the modulus of CUE characteristic polynomials}
Let us define the moments ${\cal Z}_{N}(\beta;L)$ as in (\ref{1}), and concentrate for simplicity on the case $L=2\pi$.
 Then for $0<\beta<1$ and $N\gg 1$ the probability density of the random variable $z={\cal Z}_{N}(\beta;2\pi)/ {\cal Z}_{e}(\beta,2\pi)$, where
 \be
 {\cal Z}_{e}(\beta,2\pi)=N^{1+\beta^2}\frac{G^2(1+\beta)}{G(1+2\beta)\Gamma(1-\beta^2)},
 \ee
is given by
\be \label{momdisshigh}
 {\cal P}(z)=\frac{1}{\beta^2}z^{-\left(1+\frac{1}{\beta^2}\right)}\,
e^{-z^{\frac{1}{\beta^2}}},\quad  z\ll N^{1-\beta^2}\to \infty
\ee
with $\Gamma(x)$ and $G(x)$ denoting, respectively, the Euler Gamma-function and the Barnes G- function (which satisfies
$G(x+1)=\Gamma(x)G(x), G(1)=1$).
In the region $\beta>1$ the probability density of the moments is conjectured to change to a much more complicated
distribution. Defining the scaled moments for all $\beta>1$ as $z={\cal Z}_{N}(\beta;2\pi)\,\frac{(\ln{N})^{3/2}}{N^2}$, the most salient feature of ${\cal P}(z)$  for $N\gg 1$ is predicted to be the following tail:
  \be \label{momdisslow}
 {\cal P}(z)\propto z^{-\left(1+\frac{1}{\beta}\right)}  \ln{z}, \quad z\gg 1.
\ee
Both the change of the tail exponent from $1+\frac{1}{\beta^2}$ to $1+\frac{1}{\beta}$ as well as the presence of the logarithmic factor $\ln{z}$
in (\ref{momdisslow}) are different manifestations of the freezing transition occurring at $\beta=1$. They are expected to be universal features for all values of $L$ in the range $0< L \le 2\pi$.

\subsection{Freezing of the mean Free Energy}

Perhaps the simplest consequence of freezing manifests itself in the temperature dependence of the free energy.  For ease of presentation, we again focus on the case when $L=2\pi$.  Let us define the normalized free energy by
\begin{equation}\label{f-norm}
 {\cal F}(\beta)=-\frac{1}{\beta\log N}\log{{\cal Z}_N(\beta, 2\pi)}.
\end{equation}
When $\beta$ is small, the average of ${\cal F}(\beta)$ with respect to  $U_N \in {\cal U}(N)$ is dominated as $N\rightarrow \infty$ by typical values taken by $p_N(\theta)$, those governed by the central limit theorem \cite{KS, BF}. We have seen above that the typical scale  for ${\cal Z}_N(\beta, 2\pi)$ in that case is ${\cal Z}_{e}(\beta,2\pi)=N^{1+\beta^2}$, and so we expect that
\begin{equation}
-\mathbb{E} \left\{{\cal F}(\beta)\right\}\rightarrow \left(\beta + \frac{1}{\beta}\right).
\end{equation}

One the other hand, from (\ref{2}) we know that for $\beta \rightarrow\infty$ the free energy is dominated by the extreme values taken by $p_N(\theta)$. The latter will be shown below to scale as $\max_{\theta\in[0,2\pi)}\log{|p_N(\theta)|}\approx \log{N}$.  We therefore expect that as $N\rightarrow \infty$
\begin{equation}
-\mathbb{E}\left\{{\cal F}(\beta\to \infty)\right\}\rightarrow 2.
\end{equation}

{\em Freezing} in this conext simply means that the transition between the two types of behaviour occurs at $\beta=1$ and is sharp: in the limit $N \to\infty$
\be\label{ffreezing}
-\mathbb{E}\left\{{\cal F}(\beta)\right\}=\left\{\begin{array}{c}  \beta + \frac{1}{\beta}\,\,  \quad \beta\le 1\\
2 \,\,  \quad \quad\quad \quad \beta > 1
\end{array}\right.
\ee
Specifically, the term is motivated, after interpreting $\beta^{-1}$ as the temperature $T$,  by the fact that
 the free energy remains $T-$independent, i.e.~``frozen", below the critical temperature $T=1$.

\subsection{Statistics of the maximum of the modulus of CUE characteristic polynomials}

The analogy we develop suggests that the maximal value of the modulus of a CUE characteristic polynomial $p_N(\theta)$ in an interval $\theta\in[0,L),\, 0<L\le 2\pi$
(which contains on average $N_L=N\frac{L}{2\pi}$ eigenvalues of the associated matrix $U_N$) can be written in the limit $N_L\rightarrow \infty$ as
\be\label{12sum}
-2\max_{\theta\in[0,L)}{\log{|p_N(\theta)|}}\sim a_{N_L}+b_{N_L}\,x,
\ee
where $a_{N_L}=-2\log{N_L}+c\log{\log{N_L}}+o(1)$, with, conjecturally, $c=\frac{3}{2}$ and  $b_{N_L}=1+O(1/\log{N_L})$, and where the random variable $x$  is distributed with a probability density $p(x)$. The value of $c$, $\frac{3}{2}$,  is significant because it is different from the value, $\frac{1}{2}$, characterizing the extrema of short-range correlated random variables; see the detailed discussion around equation (\ref{threshold}).
The nature of the variable $x$ and the form of the density $p(x)$ both depend on the arclength $L$, and are understood presently only for two specific choices, as detailed below.

\begin{enumerate}

\item  The  {\em full-circle} case $L=2\pi$ when $N_L=N$. Then the random  variable $x$ is of order unity and its
 probability density is predicted to be given by

\begin{equation}\label{prediction1}
p(x)=-g_{\beta_c}'(x)=-\frac{d}{dx}\left[2 e^{x/2} K_1(2 e^{x/2})\right]=2e^{x}K_0(2 e^{x/2}),
\end{equation}
where $K_{\nu}(z)$ denotes the modified Bessel function of the second kind.

The function $g_{\beta_c}(x)$ has the following expansion when $x\to -\infty$
\begin{eqnarray}\label{asympcirc}
g_{\beta_c}(x) = 1 + e^{x} ( x - 1 + 2 \gamma_E) + e^{2 x}
(\frac{1}{2} x - \frac{5}{4} +  \gamma_E) + ..\,,
\end{eqnarray}
where $\gamma_E$ is Euler's constant.

\item  A {\em mesoscopic interval} of small arclength $L\ll 2\pi$ such that  $N_L\gg 1$, so it typically contains many zeroes of the characteristic polynomial. In this case we have $x=u\sqrt{-2\ln{L}}+y$, where $u$ is a standard (mean zero, unit variance) gaussian random variable, and $y$ is independent of $u$ and is of order of unity. The probability density function for $y$ can be written again as
 $p(y)=-g_{\beta_c}'(y)$, but
$g_{\beta_c}'(y)$ is now given in terms of a contour integral:
\be\label{20sum}
g_{\beta_c}(y) = e^{y} \frac{1}{2 i \pi}
\int e^{-s y} M(s)\Gamma(s -1)\, ds
\ee
where the contour is parallel to the imaginary axis $s=s_0+ i \omega$, with $s_0$ sufficiently large that all singularities lie to the left, and $M(s)$ can be expressed in terms of the Barnes G-function $G(x)$
\be\label{21}
M(s)=\frac{2^{2 s^2 + s - 2} }{G(5/2)^2
\pi^{s-1}}  \frac{1}{\Gamma(s) \Gamma(s+2)}
\left[\frac{G(s+\frac{3}{2})}{G(s)}\right]^2.
\ee
The function can be evaluated numerically \cite{FLeDR09}, and the cumulants for the random variable $y$ can be evaluated to be
 $<y>=\mathbb{E}\{y\}=\frac{7}{2}-2\gamma_E-\ln(2 \pi)$,
$<y^2>_c=\mathbb{E}\{x^2-<x>^2\}=\frac{4 \pi^2}{3}-\frac{27}{4}$, and for general $n \geq
3$
\be\label{22}
<y^n>_c = (-)^{n-1} (n-1)! \big( \zeta(n-1) (2^n-4) - \zeta(n) (2^n 3 -4) +2^{n+1}-1-2^{-n} \big)
\ee

It is instructive to compare the behaviour of $g(y)$ at $y
\to -\infty$ with that predicted for the full circle (\ref{asympcirc}). We have
\begin{eqnarray} \label{asympint}
&& g(y) = 1 + (y + A') e^y + (A + B y +C y^2 + \frac{1}{6} y^3) e^{2 y} +\ldots
\end{eqnarray}
with $A'=2 \gamma_E + \ln(2 \pi)-1$ and $C = -0.253846$, $B =
1.25388$, $A = -5.09728$. Significantly, we see the same
asymptotic tail $g(y)-1\sim y e^y$ shared by the two functions,
but that the higher order terms in (\ref{asympcirc}) and (\ref{asympint}) differ.
The asymptotic behaviour $p(x\to -\infty)\approx -x e^x+\ldots$ is conjectured to be the
universal backward tail shared by extreme value distribution of all logarithmically-correlated random functions, see \cite{CLeD}.

\end{enumerate}

The significance of these results is that they differ from the usual Gumbel distribution\footnote{As was recently observed in \cite{KMS} the probability density (\ref{prediction1}) in fact corresponds to the sum of two independent Gumbel-distributed variables.},  which holds for maxima of a long sequence of i.i.d. random variables with finite moments (see Section 3).  This difference is essentially due to the logarithmic correlations exhibited by $V^{(U)}_N(\theta)$ and discussed in the introduction: had the correlations been short-range (see Section 3), the Gumbel distribution would have applied.  This difference shows up in the form of the distributions $p(x)$, and in particular in the asymptotic decay of $p(x)$ as $x\rightarrow-\infty$, but also, importantly, in the value of $c$ in (\ref{12sum}).  Our prediction, $c=\frac{3}{2}$ which is conjectured to be another universal feature of logarithmically-correlated processes, see  \cite{FLeDR12}, differs from that, $\frac{1}{2}$, which would be expected when the correlations are short-range (or absent).

\subsection{High points of  CUE characteristic polynomials}\label{highpoints}
It follows from (\ref{12sum}) that the typical value of the maximum of $|p_N(\theta)|$ in the interval $\theta\in[0,L]$ is of the order of $N_L$.
 The simplest quantity that quantifies the structure associated with the high values of $|p_N(\theta)|$ is
the relative length   $\mu_N(x;L)$ (as a fraction of the total length $L$) of those intervals in $[0,L]$ where $|p_N(\theta)|>N_L^x$, where $0<x<1$.
This can be expressed as
\be \label{m1}
\mu_N(x;L)=\frac{1}{L}\int_0^{L}\chi\{2\log{|p_N(\theta)|-2x\log{N_L}\}}d\theta,
\ee
where the characteristic function $\chi\{u\}=1$ if $u>0$ and zero otherwise. In the language of the theory of random processes,
quantities similar to (\ref{m1}) are known as {\it sojourn times} of the random function $2\log{|p_N(\theta)|}$ above the level $2x\log{N_L}$.
Explicit expressions for the probability density of $\mu\equiv \mu_N(x;L)$
can be provided again in the two limiting cases:
\begin{enumerate}
\item  The  full-circle case $L=2\pi$, when $N_L=N$. We denote the typical value $\mu_e(x)$ of the length $\mu_N(x;L)$ by
\be \label{m12sum}
 \mu_e(x)=N^{-x^2}\sqrt{\frac{1}{\pi\log{N}}}\frac{G^2(1+x)}{2x \, G(1+2x)}\frac{1}{\Gamma(1-x^2)}, \quad 0<x<1
\ee
The probability density for the variable $\xi=\mu_N(x;L)/\mu_e(x)$ is then predicted to have the following form:
 \be \label{m13sum}
 {\cal P}(\xi)=\frac{1}{x^2}\, \xi^{-1-\frac{1}{x^2}}\,
e^{-\xi^{-\frac{1}{x^2}}}, \quad 0<x<1
\ee

Note that the mean value of the length  $\mathbb{E}\left\{\mu_N(x)\right\}=\mu_e(x)\Gamma(1-x^2)$ stays finite as $x \to 1$.
A direct calculation shows that such an expression for the mean is valid for any
$x>0$, without restricting to $x<1$. However, when approaching $x=1$  the mean value
is significantly larger than the typical value $\mu_e(x)$, and is dominated by rare fluctuations.
This is directly related to the fact that $x=1$ is, to leading order, the typical value for the highest maximum of $|p_N(\theta)|$ (the ``extreme
value region") so the statistics of the corresponding length is indeed dominated by rare events. As will be explained, it is such a difference
between the mean and typical values which helps us to conjecture $c=3/2$ in (\ref{12sum}).

The nontrivial leading order scaling with $N^{-x^2}$ in $(\ref{m12sum})$ is directly related to the multifractal-type structure of the measure of intervals supporting high values.
The formula (\ref{m13sum}) is expected to be valid for all $\xi$ of order of unity, more precisely as long as $\xi \ll \xi_c$, with a certain cutoff scale $\xi_c\to \infty$ as $N\to \infty$, the specific form of which is, as yet, unknown to us.

\item  For a mesoscopic interval $L$ such that  $1\ll N_L\ll N$ the random variable $\mu_N(x;L)$  is distributed as the product of two statistically independent factors: $\mu_N(x;L)=e^{x\,u\sqrt{-2\ln{L}}}\tilde{\mu}_N(x)$, with the random variable $u$ being a standard mean-zero unit-variance Gaussian.
The variable  $\tilde{\mu}_N(x)$ has a typical scale $\tilde{\mu_e}(x)$  related to (\ref{m12sum}) by
$ \tilde{\mu}_e(x)=\frac{1}{(2\pi)^{x^2}}\,\mu_e(x)$, thus sharing the same multifractal scaling of the length of intervals supporting high values. The probability density ${\cal P}(\xi)$ of the random variable $\xi=\tilde{\mu}_N(x)/\tilde{\mu}_e(x)$  is expected to share the powerlaw tail ${\cal P}(\xi)\sim \xi^{-1-\frac{1}{x^2}}$ for $1\ll \xi\ll \xi_c\to \infty$ with the full-circle case, but the exact shape of the distribution will be different.

Explicitly, let us define $M_{x}(s)=\mathbb{E}\left\{\xi^{1-s}\right\}$ for complex $s$, at fixed $0<x<1$.
Then the density ${\cal P}(\xi)$ can be written as the contour integral
\be\label{FLD}
{\cal P}(\xi)=\frac{1}{\xi^2}\frac{1}{2\pi i} \int_{Re s=const}{\xi}^s\, M_{x}(s)\,ds\,.
\ee
Note that for the full-circle case $M_{x}(s)=\Gamma\left(1-x^2(1-s)\right)$, so that
performing the integral (\ref{FLD}) by the sum over residues reproduces (\ref{m13sum}).

For the mesoscopic case we have:
\begin{eqnarray} \label{MT}
&&  \! \! \! \! \! \! \! \! \! \! \! \! \! \! \! \! \! \! \! \! \! \! \! \! \! \! \! \! \! \! \! \!   M_{x}(s) = A_x(s)\frac{ \Gamma(1+ x^2 (s-1)) G_x(\frac{x}{2} + \frac{1}{x} + x s)
G_\beta(\frac{3}{2 x} + x s) G_x(\frac{x}{2} + \frac{3}{2 x} + x s) }{G_x( x + \frac{2}{x} + x s)
\left[G_x(\frac{1}{x} + x s)\right]^2} \nonumber \\
&&
\end{eqnarray}
with
\begin{eqnarray}
&& A_x(s)=2^{(s-1)(2+x^2 (2 s+1))} \pi^{1-s} \frac{\left[G_x(\frac{1}{x} + x)\right]^2 G( 2 x + \frac{2}{x})}{G_x(\frac{3 x}{2} + \frac{1}{x}) G_x(\frac{3}{2 x} + x) G_x(\frac{3 x}{2} + \frac{3}{2 x})}.
\end{eqnarray}

Here  $\Gamma_2(z|x)\equiv G_x(z)$ is the Barnes' double $\Gamma$-function \cite{Barnes}: for $\Re(z)>0$
\begin{eqnarray}\label{ZamoBarn}
 &&  \! \! \! \! \! \! \! \! \! \! \! \! \! \! \! \!  \! \! \! \! \! \! \! \! \! \! \! \! \! \! \! \! \log{G_x(z)} = \frac{x-Q/2}{2} \ln (2 \pi)\\ \nonumber &&+ \int_0^\infty \frac{dt}{t} \left( \frac{e^{- \frac{Q}{2} t} - e^{- z t}}{(1-e^{-\beta t})(1-e^{-t/\beta})}
+\frac{e^{-t}}{2} (Q/2-s)^2 + \frac{Q/2-z}{t} \right),
\end{eqnarray}
where $Q=x+1/x$. This function  satisfies:
\begin{eqnarray} \label{Gt1}
 G_x(z) = G_{1/x}(z), \quad G_x(z + x) = x^{1/2 - x z}(2 \pi)^{\frac{x-1}{2}} \Gamma(x z)\,G_x(z).
\end{eqnarray}
For $x=1$ the function  $G_x(z)$ coincides with the standard Barnes function $G(z)$ discussed after (\ref{momdisshigh}). Like the standard Barnes function, $G_x(z)$ has no poles and only zeros, and these are located at $z=-n x  - m/x$, $n,m=0,1,..$. We note in passing that
$G_x(z)$ plays a fundamental role in the Liouville model of Quantum Gravity, see e.g. \cite{FZZ}, and in recent calculations of the asymptotics of the spacing distribution at the hard edge for $\beta$-ensembles \cite{For}.
\end{enumerate}

\subsection{Extreme values of $\zeta(1/2+it)$}

The relationship between the values of $|\zeta(1/2+it)|$, as $t$ varies along the critical line, and those taken by $|p_N(\theta)|$ when the corresponding matrix $U_N$ is  chosen uniformly at random from the unitary group ${\cal U}(N)$ was first considered in \cite{KS}, where it was argued that, statistically, $\zeta(1/2+it)$ behaves like the characteristic polynomial of a random unitary matrix of dimension
$N \sim \log\frac{t}{2\pi}$.  It has since been the subject of a number of studies (see, e.g., \cite{CFKRS03, CFKRS05, GHK, CFKRS08}), but we are still far from a complete understanding.  For example, the role played by arithmetic is still being elucidated, although the hybrid model of \cite{GHK} suggests that at leading order this contribution decouples from the random matrix component.  For this reason, the way in which the random-matrix predictions listed above model the extreme value statistics of $\zeta(1/2+it)$ is not entirely clear to us.  Nevertheless, we believe that formulae at least similar to those listed below should hold at leading order.  We give our reasons for believing this in Section 5.  This belief is also supported by preliminary numerical experiments, the results of which we present below.

In the light of the understanding that  $\zeta(1/2+it)$ behaves like the characteristic polynomial of a random unitary matrix of dimension $N \sim \log\frac{t}{2\pi}$, it is natural to expect, approximately, a single matrix to model in a statistical sense the zeta function over a range $T \le t \le T+2\pi$, as such a range contains  $\log\frac{t}{2\pi}$ zeros on average.  One can thus consider splitting the critical line into ranges of length $2\pi$, and modeling each by a different unitary matrix (c.f.~\cite{FGH}).  In each range one can then find the maximum value taken by $|\zeta(1/2+it)|$, and finally one can consider the distribution of these maximum values for all of the ranges.  More generally, one can consider ranges  $T \le t \le T+L$, where $L \le 2\pi$.

Thus if
\be\label{zetamax}
\zeta_{\rm max}(L; T)=\max_{T \le t \le T+L}|\zeta(1/2+it)|,
\ee
where $0 < L \le 2\pi$, then we can anticipate that $\log \zeta_{\rm max}(L; T)$ is given by (\ref{12sum}), with $|p_N(\theta)|$ replaced by $\zeta(1/2+it)$, the maximum replaced by (\ref{zetamax}), $N_L$ replaced by $\frac{L}{2\pi}\log\frac{T}{2\pi}$, and where $p(x)$ is given approximately by the formulae above when $L=2\pi$ and when $L\ll 2\pi$, respectively.  Specifically, we expect
\be\label{zetapredict}
-2\log\zeta_{\rm max}(L; T)\sim -2\log\left(\frac{L}{2\pi}\log\frac{T}{2\pi}\right)+c\log\log\left(\frac{L}{2\pi}\log\frac{T}{2\pi}\right)+x
\ee
where the random variable $x$ has a value distribution $p(x)$ that is given, when $L \ll 2\pi$, in terms of (\ref{20sum}), and, when $L=2\pi$, that is approximated (because characteristic polynomials are $2\pi$-periodic, unlike $\zeta(1/2+it)$) by (\ref{prediction1}). This has two significant implications: that this formula holds with  $c=\frac{3}{2}$, rather than  $c=\frac{1}{2}$, as would be the case if the zeta correlations were short-range, and that the tail of the distribution decays like $|x|e^x$ as $x \rightarrow -\infty$.  It is not at this stage completely clear to us how, if at all, the arithmetic will modify these expressions, but there are reasons, discussed in Section 5, to believe that it will not influence them at leading order.

Furthermore, we expect, with the same identifications,
\be\label{zetameas}
\mu_T(x;L)=\frac{1}{L}{\rm meas}\left[T\le t\le T+L: 2\log|\zeta(1/2+it)|\ge 2x\log\left(\frac{L}{2\pi}\log\frac{T}{2\pi}\right)\right]
\ee
to be given by the corresponding expressions listed above.  In this case we do expect the scale (\ref{m12sum}) to be multiplied by the arithmetical factor
\be\label{arith}
a(x)=\prod_p \left[\left(1-\frac{1}{p}\right)^{x^2}
\sum_{m=0}^\infty\left(\frac{\Gamma(x+m)}{m!\Gamma(x)}\right)^2p^{-m}\right]
\ee
that appears in the moment conjectures at leading order \cite{KS}.  Specifically, we expect the value distribution of $\mu_T(x;L)$ to be given by the formulae in Section 2.3, but with $\tilde{\mu}_e(x)$ replaced by
\be
a(x)\left(\log\frac{T}{2\pi}\right)^{-x^2}\sqrt{\frac{1}{\pi\log\log\frac{T}{2\pi}}}\frac{G^2(1+x)}{2x \, G(1+2x)}\frac{1}{\Gamma(1-x^2)}.
\ee

Finally, we expect to see freezing of the quantity corresponding to the free energy; that is, defining
\be
{\cal Z}_T(\beta)=\frac{1}{2\pi}\log\frac{T}{2\pi}\int_T^{T+2\pi}|\zeta (1/2+it)|^{2\beta}dt
\ee
and
\begin{equation}\label{f-norm}
{\hat {\cal F}}_{\zeta}(\beta)=-\frac{1}{\beta\log\log\frac{T}{2\pi}}\log{{\cal Z}_T(\beta)},
\end{equation}
then we expect that the mean of ${\hat {\cal F}}_{\zeta}(\beta)$ with respect to $T$ satisfies
\be\label{ffreezingzeta}
-\left<{\hat {\cal F}}_{\zeta}(\beta)\right>=\left\{\begin{array}{c}  \left(\beta + \frac{1}{\beta}\right)\,\,  \quad \beta\le 1\\
2 \,\,  \quad \quad\quad \quad \beta > 1
\end{array}\right.
\ee
in the limit as $T\rightarrow\infty$.

It is worth remarking that our results relate to extreme values over much shorter ranges than those considered by Farmer, Gonek $\&$ Hughes \cite{FGH} - our focus is on ranges of lengths that are $O(1)$, whereas theirs was on ranges of length $T$ as $T\rightarrow \infty$.  If one extrapolates (\ref{12sum}) to ranges of length $T$ (well beyond where we can justify it), our result for the typical scale of the extreme values agrees with theirs.  Where we
are able to go further is in predicting the value distribution of the fluctuations.  In the context of the question of the extreme values for $t<T$, it is worth noting again that the tail of the distribution we predict for much shorter ranges decays like $|x|e^x$ as $x \rightarrow -\infty$; that is, the exponential is linear rather than quadratic.  If this were to persist to much longer ranges than we understand at present, it would suggest that  $\zeta(1/2+it)$ may take much larger values than the Montgomery heuristic (or, more precisely, the Farmer-Gonek-Hughes conjecture)  predicts, maybe even close to the upper limit (\ref{Ozeta}), but there are several reasons for thinking this unlikely.  Specifically, it is not difficult to see that in the absence of correlations the large deviation tail of the extreme value density again becomes quadratic, and we believe the same is also true when log-correlations are present.

\subsection{Numerical experiments}

In order to test the extreme value predictions for the Riemann
zeta function we now summarize the results of preliminary numerical computations performed by Dr Ghaith Hiary and published in our short communication FHK.  These involved evaluating $\zeta(1/2+it)$ over ranges of length $2\pi$, at various heights $T$, and finding the maximum value $\zeta_{\rm max}(2\pi; T)$ in each range.  Values of $\zeta(1/2+it)$ were computed
using the amortized-complexity algorithm of \cite{H}, which is suitable for computing $\zeta(1/2 + it)$ at many points.
Point-wise values of $\zeta(1/2+it)$ that were computed are typically accurate to within $\pm 5\times 10^{-11}$, which is sufficient for the purposes of this experiment. The maximum of $|\zeta(1/2 + it)|$ between consecutive zeros was computed to within $\pm 10^{-9}$.

The first test concerns the value of the constant $c$ in  (\ref{zetapredict}).  We expect the logarithmic correlations to lead to $c=\frac{3}{2}$, rather than  $c=\frac{1}{2}$, as would be the case if the zeta correlations were short-range. The mean of $\zeta_{\rm max}(2\pi; T)$ suggested by the model in (\ref{zetapredict}) is $\delta = e^{\gamma_E} N/(\log N)^{\frac{c}{2}}$, with $c = 1/2 \textrm{ or } 3/2\,,$ and $\gamma_E = 0.57721\ldots\,,$ where we set $N$ to be the nearest integer to $\log T$.
At each height a sample that spans $\approx 10^7$ zeros is used yielding $\approx 10^7/N$ sample points
(since there are roughly $N$ zeros in each range of length $2\pi$).
The numerics presented in FHK and reproduced below clearly shows that $c = 3/2$ fits the data considerably better than $c = 1/2$, thus supporting the
logarithmic correlations model.
\begin{table}[ht]
\footnotesize
\caption{\footnotesize Ratio of data mean $\tilde{\delta}$ to model mean $\delta$ with $c = 3/2$ and $c = 1/2$.}\label{mean ratios}
\renewcommand\arraystretch{1.5}
\begin{tabular}{c|c|c|c}
$T$       & $ \qquad N \qquad $        &   $\left(\tilde{\delta}/ \delta\right)_{c=3/2}$ & $\left(\tilde{\delta}/ \delta\right)_{c=1/2}$\\
\hline
$10^{22}$ & 51  &     1.001343    &   0.504993\\
$10^{19}$ & 44  &    0.992672   & 0.510293    \\
$10^{15}$ & 35  &    0.976830    & 0.518057    \\
$3.6\times 10^{7}$ & 17  &  0.930533 &  0.552856
\end{tabular}
\end{table}

Testing the distribution $p(x)$ is more difficult, because the data converge extremely slowly at that scale.  The results of some initial experiments were presented in FHK and are reproduced below.  Specifically, we considered
\be
\sigma(T)=-2 \log |\zeta_{\rm max}(2\pi; T)| + 2\log\log\frac{T}{2\pi} - \frac{3}{2} \log\log\log\frac{T}{2\pi}
\ee
based on a set of approximately $2.5 \times 10^8$ zeros
near $T = 10^{28}$. The data were normalized so that $\sigma(T)$ has empirical variance
$= \int x^2 p(x)\,dx =3.28986813\ldots$.  The overall agreement was supportive of (\ref{prediction1}), especially in the important tail when $x\to -\infty$ and in view of the fact that lower order arithmetical terms \cite{KS, CFKRS05} had not been incorporated, but it cannot be said to be conclusive at this stage.  The behaviour in the tail is significant because if it were to persist into the large deviation regime it would suggest that the Montgomery heuristic (or, more precisely, the Farmer-Gonek-Hughes conjecture) significantly underestimates the maximum values achieved by $|\zeta(1/2+it)|$, however, as noted above, there are strong reasons for believing that it does not persist this far.

\begin{figure}[h!]
\begin{center}
\includegraphics[width=0.8\textwidth, height=0.5\textheight]{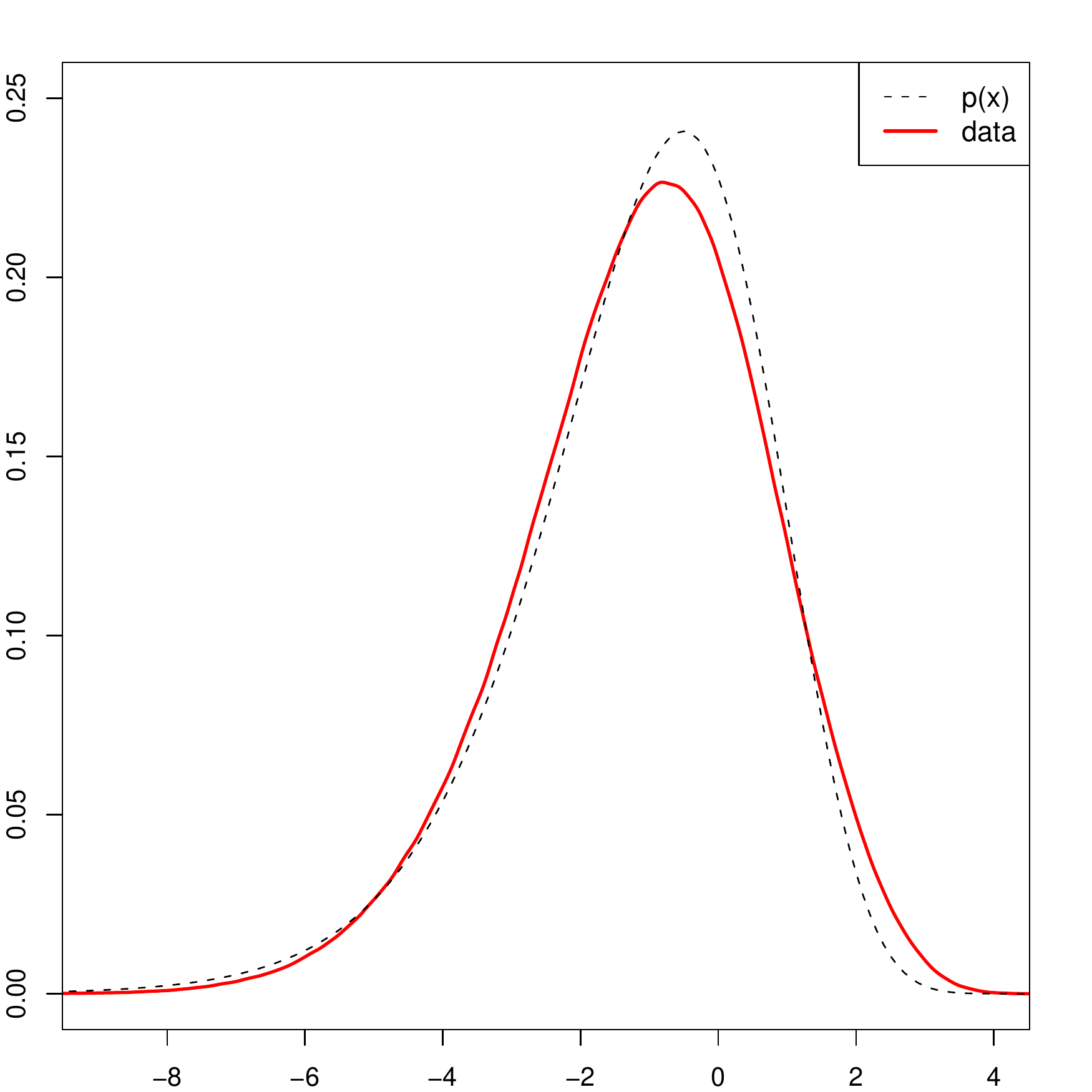}
\end{center}
\caption{Numerical computation (solid red line) compared to theoretical prediction (\ref{prediction1}) (dashed black line) for $p(x)$.}
\label{figure 1}
\end{figure}

\begin{figure}[h!]
{\small }
\begin{center}
\includegraphics[width=0.8\textwidth, height=.5\textheight]{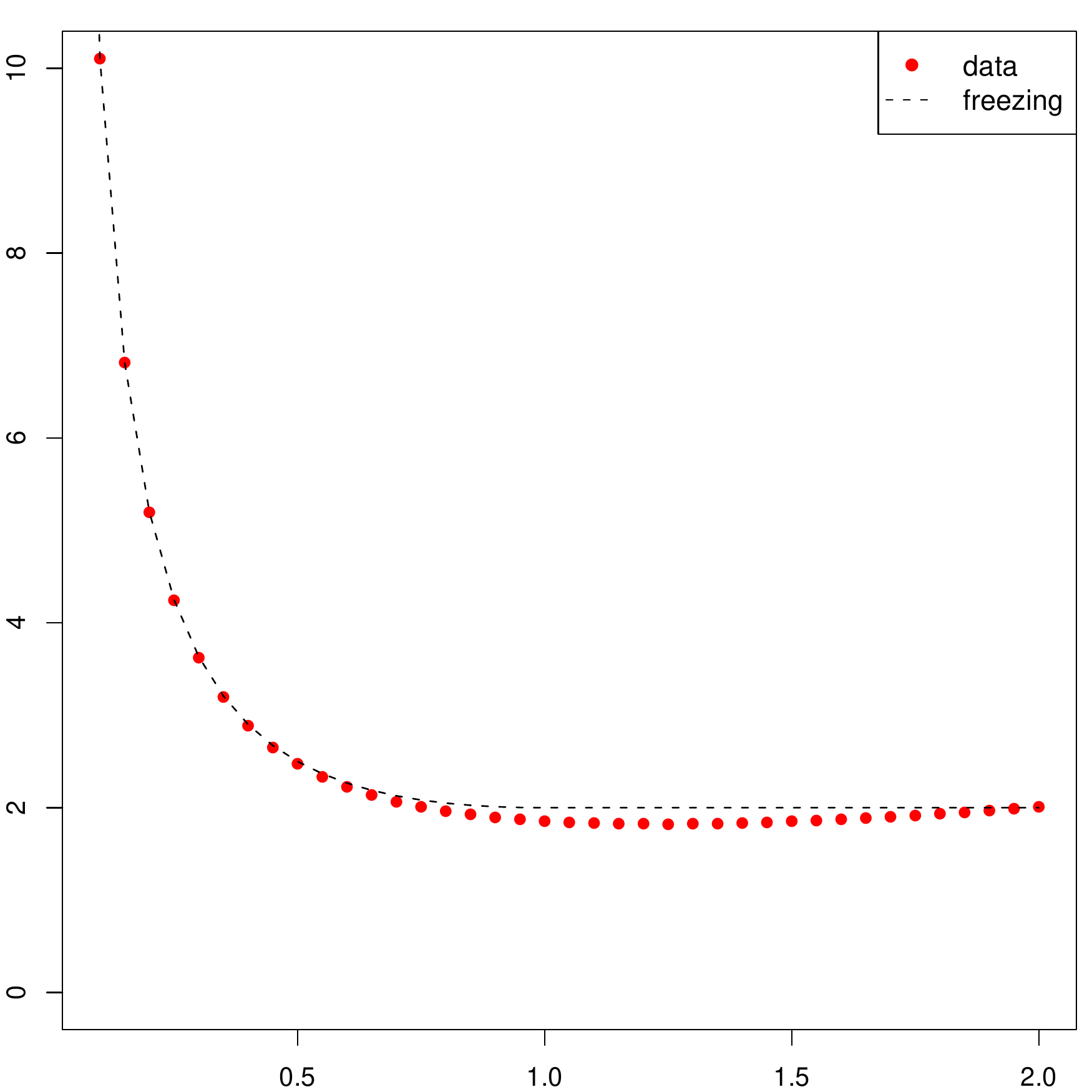}
\end{center}
\caption{Numerical computation (red dots) compared to the theoretical prediction (dashed black line) for  $D_T(\beta)$,
suggesting freezing beyond $\beta=1$}\label{figure 2}
\end{figure}

Finally, we performed a test of the prediction relating to freezing in the mean free energy.  This involved calculating ${\cal Z}_T(\beta)$ and hence the free energy ${\hat {\cal F}}_{\zeta}(\beta)$ numerically, and then averaging with respect to $T$  over $10^{6}$  values near to $T=10^{28}$.  If freezing is operative, $-{\hat {\cal F}}_{\zeta}(\beta)$  is expected to be equal to $\beta+1/\beta$ for $\beta<\beta_c=1$ and remain frozen to $-{\hat {\cal F}}_{\zeta}(\beta)=2$
for all $\beta>1$.  In order to account for the finite height at which the computations could be carried out, it was found to be efficacious to normalize ${\cal Z}_T(\beta)$ so as to incorporate lower terms in the full asymptotic expansion of the moments \cite{CFKRS05, CFKRS08} rather than just the leading-order asymptotic term.  Specifically, what was computed was the $T$-average of
\be
D_T(\beta)=\beta+\frac{1}{\beta}+\frac{1}{\beta\log\log\frac{T}{2\pi}}\log\left(\frac{{\cal Z}_T(\beta)}{\log\frac{T}{2\pi}P_{\beta}(\log\frac{T}{2\pi})}\right),
\ee
where $P_{\beta}(x)$ denotes the moment polynomial considered in \cite{CFKRS05, CFKRS08} (and its extension as an infinite series when $\beta$ takes non-integer values).
The results shown in Figure \ref{figure 2} would appear to support freezing.

\begin{figure}[h!]
\begin{center}
\includegraphics[width=0.8\textwidth, height=0.5\textheight]{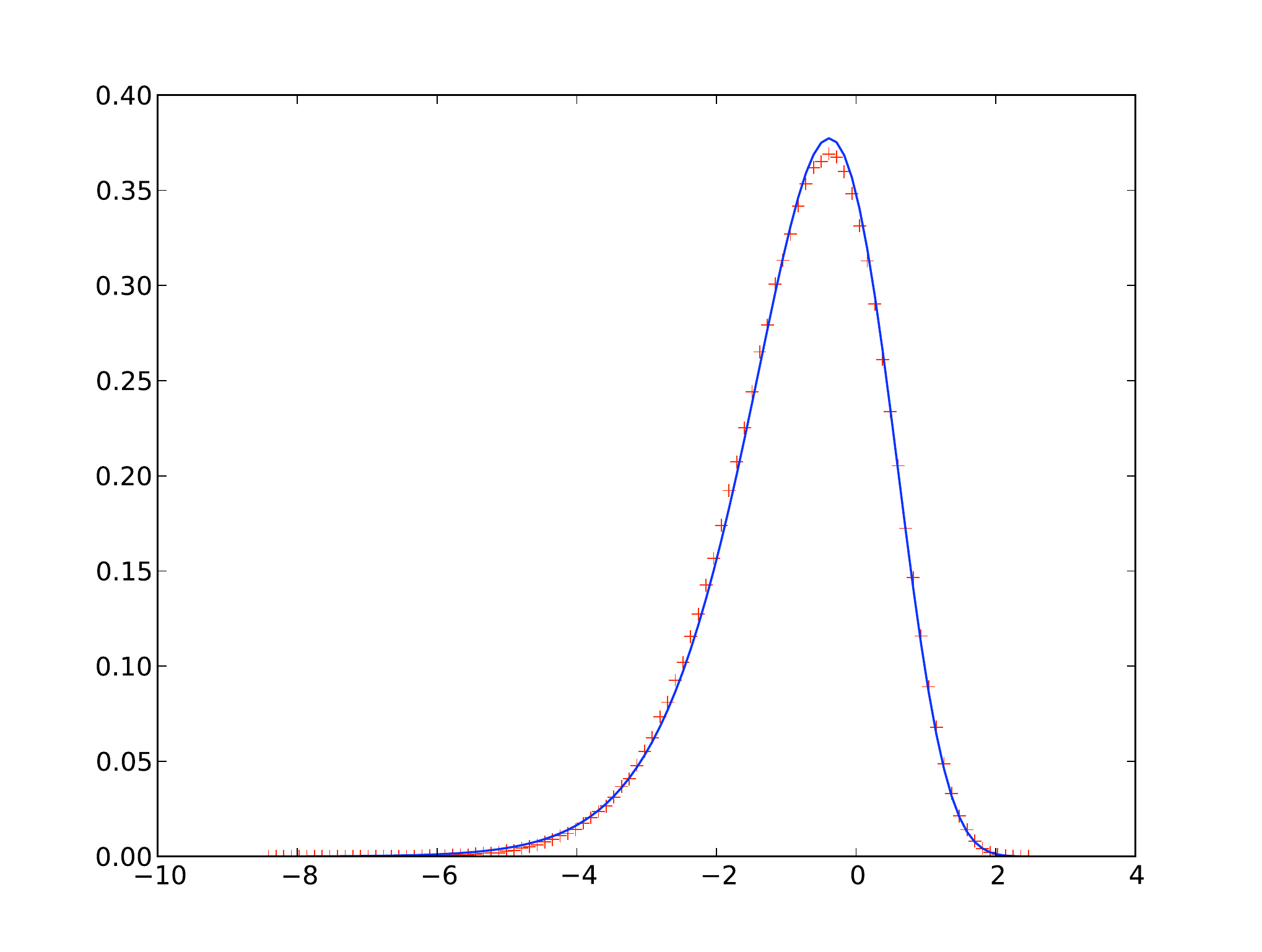}
\end{center}
\caption{Numerical computation (red crosses) for $10^6$ matrices with $n=50$ compared to theoretical prediction (\ref{prediction1}) (blue line) for $p(x)$.}\label{figure CUE}
\end{figure}

Mr Timoth\'{e}e Wintz also assisted us by performing similar numerical experiments on randomly generated unitary matrices.  Specifically, he computed the maximum values of characteristic polynomials of large numbers of unitary matrices drawn uniformly
from the CUE.  He tested the value of $c$ in (\ref{12sum}) with the results set out in Table \ref{rmt}, obtained as for the zeta-function.  There is again good agreement with our conjectured value $c=\frac{3}{2}$.  He also tested the distribution of the random variable $x$.  In this case it proved efficacious to rescale $x$ by a factor $b_N=1-B/\log N$, cf.~(\ref{12sum}),  where the constant $B$ was obtained from a best fit.  The results for a sample of a million matrices with $N=50$ are shown in Figure \ref{figure CUE}.

\begin{table}[ht]
\footnotesize
\caption{\footnotesize Numerical estimation of $c$ in (\ref{12sum}) for matrices of size $N$.}\label{rmt}
\renewcommand\arraystretch{1.5}
\begin{tabular}{c|c}
$N$        &   $c$\\
\hline
20 & 1.43570\\
30 & 1.46107    \\
40 & 1.48018    \\
50 & 1.49072\\
60 & 1.49890\\
70 & 1.50756
\end{tabular}
\end{table}

\section{Statistical Mechanics in Disordered Landscapes and Extreme Value Statistics}

The idea of complicated energy landscapes
pervades the theoretical description of glasses, disordered
systems, proteins, etc.~\cite{Wales}, and has recently re-emerged in string
theory and cosmology, see e.g.~\cite{Suss}.  In
this respect, the Parisi solution for mean-field spin-glasses is especially important:
it reveals that in that case the energy landscape of a system of many randomly interacting spins
  has a surprisingly complex, hierarchical structure of valleys within
valleys within valleys (for a short recent account see \cite{Paris} and references therein).
Such a structure manifests itself in both dynamics and thermodynamics via a non-trivial phase transition occuring at some finite
temperature $T_c>0$. Below $T_c$, dynamics associated with wandering in this maze of valleys
is non-ergodic and shows many distinguished features like aging \cite{Bouch}.
Such features are commonly observed in real experiments, although the extent to which the Parisi
theory describes energy landscapes typical for finite-dimensional disordered systems is still a matter of debate.

Investigating the energy landscape of a real interacting disordered or complex system is a notoriously difficult problem, and
an important role is played by studying effective single-particle counterparts. Here the main goal
is to describe the behaviour of the whole complex system, or
one of its subparts, by focussing on the statistical mechanics
of a single point particle (or sometimes higher dimensional
objects like lines or membranes) moving in a random
potential, which encodes the complexity of the original system.
The hope then is to be able to classify the possible types of
landscapes and to establish generic, universal properties, not unlike those emerging in Random Matrix Theory.
The most famous models of this type are the Random Energy Model (REM or GREM \cite{Derr81, Derr85}) and its later ramification describing
 the model of a polymer on a tree with a disordered potential \cite{DS}.

Recall that the equilibrium statistical mechanics for a system characterized  by a (discrete) set of energies $E_1,\ldots, E_M$  and a temperature $T>0$ is represented by the set of Boltzman-Gibbs probability weights, $p_1,\ldots, p_M$, where
\be\label{Gibbs}
p_i=\frac{1}{Z(\beta)}e^{-\beta E_i}, \quad {Z}(\beta)=\sum_{i=1}^M\,e^{-\beta E_i},
\ee
$\beta=1/T$, and we set the Boltzmann constant $k_B$ to unity to measure the temperature and the energy in the same units.
 It is clear that at low temperatures, $\beta\gg 1$, the set of probabilities is dominated by
 the lowest available energies in the set. It is then not surprising that one of the most fundamental questions arising in the landscape paradigm is the problem of  understanding the statistical properties of low, or even ``extreme" (i.e. minimal)
 energy values typical for various classes of disordered landscapes \cite{BM1, BBP}.
 Such an understanding is certainly needed for a detailed description of the freezing
phenomena in systems with disorder, with the spin-glass-like arrest being the paradigmatic example.
From that angle, the analysis of the statistics of minimal-energy configurations of various random systems is
attracting a good deal of attention at present; for a review of some recent developments related to Tracy-Widom
type statistics see \cite{Dots}.

In mathematics, the distribution of the minimum/maximum $V_{min}$  in a sequence of $M\gg 1$ random variables $V_1,\ldots,V_M$
is an important research area with numerous applications.  The classical results in this area can be
found, for example, in \cite{LLR}, and we attempt to summarize the facts most pertinent to our present study below.
  If $V_i$ are i.i.d. random variables there are only three possible shapes (up to shifts and rescaling) of the limiting distribution $\Phi(V)=Prob(V_{min}>V)$ of the minimum $V_{min}$.  In particular, for i.i.d. variables $V_i$ with all moments finite
   the relevant distribution has a characteristic double-exponential form and is known as the Gumbel distribution. More precisely, there exist non-random sequences $a_M$ and $b_M$ such that the random variable $y=(V_{min}-a_M)/b_M$ is characterised when $M\to \infty$ by the limiting distribution
 $\Phi(y)=\exp\{-e^{y}\}$. In the particular case of i.i.d.~normal,  mean zero variables $V_i$  with variance
 $\mathbb{E}\{V_i^2\}=\sigma<\infty$  the sequences $a_M,b_M$ behave asymptotically like $a_M\approx -\sqrt{2\sigma \log{M}},\, b_M\approx \sqrt{{\sigma}/{(2\log{M})}}$.

An important general question, relevant for applications, is to what extent, if at all, the above picture holds for correlated random sequences.
The most complete answer is known for Gaussian mean-zero stationary sequences with covariance $\mathbb{E}\{V_iV_j\}=C(|i-j|)$. It turns out that if $C(r)$ decays to zero faster than $1/\log{r}$ the limiting distribution of the minima
is still given by the Gumbel distribution, and the leading-order scaling behaviour for $a_M,b_M$ is the same as for uncorrelated sequences.
We will call such variables {\em short-range correlated}.  Not much is known at present beyond the short-range correlated case.  There has been particular interest in
 {\it scale-invariant} sequences with  stationary increments, also known as Fractional Random Walks, which have numerous applications.
Such sequences are conveniently characterized by structure functions $\mathbb{E}\{(V_i-V_j)^2\}\propto |i-j|^{2H}$, with parameter $0<H<1$ known as the Hurst exponent. In particular, scale-invariance implies that
 the typical minimum in such a case should scale for large $M$ as $V_{min}\approx - C(H) \, M^H$ which should be contrasted with
 the short-range scaling $V_{min}\approx -\sqrt{2\sigma \log{M}}$. The probability distribution of the minimum is known explicitly only for the case
 of the standard diffusive random walk $H=1/2$ \cite{MC}, when one can exploit
 path-integral methods based on  the underlying Markovian structure.  Characterizing extreme value statistics for non-Markovian random walks $H\ne 1/2$ remains a considerable challenge; even the constant $C(H)$ is not yet known explicitly beyond $H=1/2$.

\subsection{Random Energy Model and the two-dimensional Gaussian Free Field}

The results reviewed above can already be used to generate some insight into the
 equilibrium statistical mechanics of a single particle in a disordered landscape.
 To this end let us identify $E_i\leftrightarrow V_i$ and look at the sequence
 $V_i$ as representing a set of energies available for a particle at various ``sites" $i=1,\ldots,M$ of a disordered system.
 Recall that  the free energy $ F(\beta)=-\beta^{-1}\log{{Z}(\beta)}$ can be represented as ${F}(\beta)=U-TS$, where $U=\overline{(E_i)}_T$ is the mean energy of the system,
 with $\overline{(...)}_T$ standing for the thermal average with respect to the Boltzmann-Gibbs measure (\ref{Gibbs}). The quantity $S$ stands for the
 {\it entropy}, which effectively controls how the available mean energy is spread over all the available energy levels (i.e. over the {\em sites}). In particular, the entropy of a system in which all $M$ states are equally likely is given, according to the
Boltzmann formula, by $S=\log{M}$.
   One may then attempt to understand the structure of a Boltzmann-Gibbs measure at a crude qualitative level by invoking
     the argument of a competition between the entropic term $TS=T\log{M}$ and the minimal available energy \cite{CLeD}. Recalling that $V_{min}\approx -\sqrt{2\sigma \log{M}}$ for short-ranged random sequences, we immediately conclude
that for any fixed $T>0$ the entropic contribution will dominate over the energetic component in sufficiently large systems. This should result
in a Boltzman-Gibbs measure being spread more or less  democratically over all the sites of the landscape.
The normalisation $\sum_{i=1}^M p_i=1$ then implies the scaling $p_i\sim M^{-1}$. In such a situation it is conventional to say that for any $T>0$ the system stays in the {\it high-temperature} phase with a delocalised  Boltzman-Gibbs measure. For the long-range correlated case, however, the situation is somewhat the opposite: for any temperature $T<\infty$ and Hurst exponent $H>0$ the magnitude of the minimal energy for large enough $M$ grows faster than the entropic contribution. In this situation the Boltzmann-Gibbs measure for large enough systems  will be essentially {\em localised} on one or a few sites of minimal
energy, with the corresponding $p_i= O(1)$, whereas for the majority of the sites $p_i$ is expected to be negligible. It is conventional to say that effectively such a system is {\em frozen} in the low-temperature phase, for all temperatures.

It is possible to augment the picture just described in several ways.
 The simplest is to {\it rescale} the variance $\sigma$ of the short-ranged random potential  with $M$
 in such a way that $\sigma\sim \log{M}$. We can thus ensure that both the minimal energy and the entropy grow logarithmically with $M$, and therefore
 which of the two dominates will depend on the temperature $T$. If in addition we assume the random energies to be i.i.d. random variables, the resulting model is
 precisely the much-studied Random Energy Model (REM), introduced by Derrida \cite{Derr81,Derr85}, which is sufficiently simple to be amenable to rigorous analysis and has played a paradigmatic role in the statistical mechanics of disordered systems. In particular,
  the model displays a nontrivial ``freezing" phase transition at a finite temperature $T_c$, which by an appropriate choice of the
  variance can be made equal to unity.  Namely, the mean free energy of the system
   in the limit $M \rightarrow\infty$ behaves in the high-temperature phase $T>T_c=1$ as $\mathbb{E}\{F(\beta)\}\approx -\left(\beta+\frac{1}{\beta}\right)\log{M}$ and ``freezes" to the minimal value  $\mathbb{E}\{F(\beta)\}\approx - 2\log{M}$ for all temperatures below the transition, for $0\le T\le 1$.

  In fact, both above and below $T_c$ the Boltzmann-Gibbs probability measure is neither truly localised nor delocalised, but rather provides the simplest example of a random {\it multifractal} measure.
  Namely, in the thermodynamic limit $M\to \infty$ the weights $p_i$ scale differently on different sites:  $p_i\sim M^{-\alpha_i}$ ,  with the exponents $\alpha_i$ filling some interval $[\alpha_{-},\alpha_{+}$] in such a way that $\sum_i\delta(\alpha-\alpha_i)\sim M^{f(\alpha)}$, with a well-defined smooth concave function $f(\alpha)$, see e.g.~\cite{Fyod10, FLeDR12} for a more extensive discussion and further references. A detailed analysis of the low-temperature phase reveals even more intricate probabilistic structure: the weights $p_i$ for $T<T_c$
 can be described in terms of Ruelle probability cascades \cite{Bov}. Physicists usually refer to the low-temperature structure arising as one reflecting the simplest nontrivial mechanism of a spin-glass type phase transition - the so-called one-step spontaneous replica symmetry breaking. A ramified version of the same mechanism in more sophisticated spin glass models gives rise to the Parisi picture of hierarchical valleys for the effective (free) energy landscape mentioned above.

Despite the success of Derrida's idea to rescale the variance of the i.i.d.~variables with $\log{M}$, inspired by a similar scaling in infinite-dimensional mean-field spin-glass models, such a procedure for generating a nontrivial freezing transition
looks somewhat artificial from the point of view of random landscapes over finite-dimensional lattices.
It is therefore worth noting that a similar scaling of the variance arises naturally
in certain finite-dimensional models of physical interest.  It was revealed in \cite{CMW} how the corresponding landscape model emerges
in the case of a quantum Dirac particle moving in the two-dimensional plane subject to a transverse random magnetic field.
 Specifically, it was noted there that the profile of a (normalized) eigenfunction corresponding to zero energy of the Dirac Hamiltonian has, formally, the shape of ( a continuous space version of) the Boltzmann-Gibbs measure on ${\bf x}=(x,y)\in \mathbb{R}^2$, with the role of the random potential $V({\bf x})$ played by the two-dimensional Gaussian Free Field (2dGFF). The latter is defined as a mean-zero random Gaussian field
 in a domain $D\in \mathbb{R}^2$ such that its covariance is given by  $\mathbb{E}\{V({\bf x})V({\bf x'})\}=G({\bf x},{\bf x}')$, where $G({\bf x},{\bf x}')$ is the Green function of the Laplace operator $\Delta=\partial_x^2+\partial_y^2$ in $D$, with specified conditions on the boundary
$\partial D$. Let us consider,  for definiteness, Dirichlet boundary conditions and take $D$  to be a two-dimensional disk $|z|\le L$, where we have employed the complex coordinate $z=x+iy$. The corresponding Green function is then  $G(z_1,z_2)=-\frac{1}{2\pi}\log{\frac{L|z_1-z_2|}{L^2-z_1z_2}}$.   In particular, for any two distinct points $z_1$ and $z_2$ well inside the disk $|z_{1,2}|\ll L$  we
find $G(z_1,z_2)=-\frac{1}{2\pi}\log{\frac{|z_1-z_2|}{L}}$, the latter expression interpreted as the Green function on the full two-dimensional plane.
To make the construction well-defined from the point of view of an  underlying random Gaussian field,
 one has to ensure its variance is finite by taking appropriate care of the divergence when $z_1\to z_2$. The most natural way is to think of an underlying lattice structure, with
the lattice spacing $a\ll L$ and the random field defined on the lattice sites only\footnote{It is possible to give a {\it bona fide} mathematical construction of the continuous 2dGFF, see  \cite{Sheff}, but for our goals it suffices to rely upon this heuristic approach.}. We have $M\sim (L/a)^2$ lattice points inside the disc, and it is consistent to require $\mathbb{E}\{V({\bf x})V({\bf x'})\}|_{{\bf x}\to {\bf x'}}= -\frac{1}{2\pi}\log{\frac{a}{L}}\propto \log{M}$ showing that the 2dGFF indeed naturally gives rise to a REM-like scaling of the variance. In contrast to the REM landscape, however, the values of the 2dGFF at different lattice points
are strongly (logarithmically) correlated. Nevertheless it is natural to conjecture that the freezing transition typical for the REM
(and shared by the model of directed polymers on disordered trees \cite{DS}) will also occur in such a situation, and will give rise to
a multifractal structure in the associated Boltzmann-Gibbs measure, and hence for the zero-energy wavefunction of the Dirac particle.
Numerical simulations and further analytical studies confirm the validity of such a picture, see  \cite{CLeD}. This has provided significant insight into the associated statistics of the minima of the regularized 2dGFF landscape.
In particular, it suggested a powerful, albeit heuristic, real-space renormalization group approach. This approach substantiated the claim
of a REM-like freezing scenario in logarithmically correlated landscapes, and led to a conjecture
for the minimum value of such a landscape:  $V_{min} = a_M+b_M\, y$, where
 $a_M= (- 2 \ln M + \frac{3}{2} \ln \ln M +o(1))$, $b_M=1 +O(1/\ln(M))$ and the value distribution of the random variable $y$ is given asymptoticially, as $y \rightarrow\infty$, by $\Phi(y)\sim 1-|y|e^{y}$, and so is different from the Gumbel expression, for which $\Phi(y)\sim 1-e^{y}$.

In the mathematical literature the probabilistic properties of the extremes of the lattice version of 2DGFF have attracted considerable
attention \cite{BDG, Dav, BZ}.  Actually, these papers addressed $V_{max}$ rather than $V_{min}$, but the two statistics are obviously trivially related. In particular, Bramson $\&$ Zeitouni \cite{BZ} proved that $V_{max}=2\log{M}-\frac{3}{2}\ln{\ln{M}}+O(1)$.  This agrees with the leading order terms in $a_M$. A more detailed characterisation of the extreme value distribution was beyond the reach of the methods used until very recently\footnote{See recent progress in
\cite{BDZ}.}, but the work of Davioud \cite{Dav} provided key insights into the very high, {\em almost extreme} values of the 2dGFF.
Specifically, let us  define  $N_M^{+}(x)$, where $x\in(0,1)$, to be the number of  lattice points such that the potential
$V_i$ satisfies $V_i>2x\log{M}$. It turns out that $\lim_{M\to \infty}\frac{\mathbb{E}\{\ln{N_M^{+}(x)}\}}{\ln{M}}=1-x^2$. This implies that  the typical value of $N_M^{+}(x)$ scales in every realisation roughly as $N_M^{+}(x)\sim M^{1-x^2}$. Such a dependence is natural to interpret again as a kind of multifractal scaling, not unrelated to the multifractality of the Boltzmann-Gibbs weights. Note that when $x\to 1$ the typical number of points above such a level becomes of the order of unity, which agrees with the scaling of the extreme values described above. We will return to these issues later on in the paper.

Finally, we mention that a mathematically rigorous framework for dealing with general fields and processes with logarithmic correlations was developed in \cite{Kah} and is known by the name "Gaussian Multiplicative Chaos". This has undergone substantial development  in recent years, see e.g.~\cite{BM2}, and has in particular been exploited in the context of probabilistic aspects of
 Quantum Gravity, see e.g.~\cite{DRSV} and references therein. Very recently such a framework was shown to be of substantial utility
 also for studying Boltzmann-Gibbs measures associated with $1/f$ noise landscapes, and closely related problems \cite{AZ, AJKS, BKNSW}.

\subsection{1/f-noise as a random landscape I: Statistical Mechanics in the high-temperature phase}

To attack the problem from a different angle, the idea was proposed in \cite{FB08, FLeDR09} to use the full-plane logarithmic GFF  to construct various {\it one-dimensional} Gaussian random landscapes with logarithmic correlations. The associated Boltzmann-Gibbs measures are expected to be qualitatively analogous to those in 2dGFF landscapes, but are amenable to much more detailed quantitative analysis. Arguably the simplest example of such a one-dimensional landscape can be generated by sampling the values of the full-plane 2dGFF along a circle of unit radius parametrized as $z=e^{it}, \, t\in[0,2\pi)$. One is thus led to a $2\pi-$periodic mean zero Gaussian process $V(t)$  whose covariance for $t_1\ne t_2$ is formally given by
 \be \label{logcovarcont} \mathbb{E} \{V(t_1)V(t_2)\}=-\frac{1}{2\pi}\log{|e^{it_1}-e^{it_2}|}=-\frac{1}{\pi}\log{2|\sin{\frac{1}{2}(t_1-t_2)|}}.
 \ee
  By employing the well-known identity:
$-\log{\left(4\sin^2\frac{t_1-t_2}{2}\right)}=2\sum_{n=1}^{\infty}\frac{1}{n}\cos{n(t_1-t_2)}$
 we see that one representation of $V(t)$ is
given by a random Fourier series of the form
 \be \label{1/f}
 V(t)=\sum_{n=1}^{\infty}\frac{1}{\sqrt{\pi n}} \left[v_n e^{i n t}+\overline{v}_n e^{-i n t}\right]\,,
\ee
  where the coefficients $v_n,\overline{v}_n$, $n=1,2\ldots$, are i.i.d., mean-zero, complex Gaussian variables with variance  $ \mathbb{E} \{v_n \overline{v}_n \}=1$. In this way we sample every Fourier series  (\ref{1/f}) according to the Gaussian probability weight $\prod_{n=1}^{\infty}\left[e^{-\frac{1}{2} |v_n|^2}\, \frac{dv_n\overline{dv_n}}{2\pi}\right]$.
  As the power associated with a given Fourier harmonic with index  $n$  decays like $1/n$, the function $V(t)$,
   viewed as a time-dependent random signal, is a representative of the so-called $1/f$ noise believed to be ubiquitous in Nature, see e.g.~\cite{Mil}. In the present section we shall focus on this particular model, although later on in the paper we
   will also consider another model of  $1/f$ noise, which corresponds to sampling the values of the full-plane 2dGFF along a finite interval of a straight line.

   Pretending for the moment that (\ref{1/f}) defines a well-behaving function and further exploiting the identity
   \be\label{1g}
   \fl -\frac{1}{2\pi^2}\int_0^{2\pi}\int_0^{2\pi}\ln{|2\sin\left(\frac{t_1-t_2}{2}\right)|} e^{int_1}e^{-imt_2}\,dt_1\,dt_2=\frac{1}{|n|}\delta_{n,m},\quad n \ne 0, m\ne 0
   \ee
  we see that in the space of functions $V(t)$ defined by (\ref{1/f}) we should have
   \be\label{1h}
 \fl {\cal S}(V)= -\frac{1}{2\pi}\int_0^{2\pi}\int_0^{2\pi} \ln{|2\sin\left(\frac{t_1-t_2}{2}\right)|}\,V'(t_1)\, V'(t_2)\,dt_1\,\,dt_2=\sum_{n=1}^{\infty}|v_n|^2>0.
   \ee
 where we have introduced the formal derivative $V'(t)=\frac{d V(t)}{dt}$.  The relation (\ref{1h}) then implies that the Gaussian weight
 associated with every Fourier series (\ref{1/f}) is proportional to $e^{-\frac{1}{2}{\cal S}(V)}$ so that
 the expected value for any functional $\Phi(V)$ of $1/f$ noise could be written symbolically as a formal ``path integral"
   \be\label{1i}
  \mathbb{E}\left\{\Phi(V)\right\}=\int e^{-\frac{1}{2}{\cal S}(V)} \Phi(V)\,{\cal D}V
  \ee
  much in the same way as the expected value of a functional $\Phi(W)$ of the standard Brownian motion $W(t)$
     with respect to the Wiener measure can be symbolically written as $\int e^{-\frac{1}{2} S(W)} \Phi(W)\,{\cal D}W,$ with  $S(W)=\int [W'(t)]^2\,dt$, see e.g.~\cite{Klaud} for historical remarks and further references. The crucial  difference  however is that in the case of the Wiener measure, the evaluation of expected values in many practically interesting cases can be rigorously performed by the Feynman-Kac formula, reducing calculations to the solution of a differential equation, whereas a mathematically rigorous theory behind (\ref{1h})-(\ref{1i}) is not yet available\footnote{See, however, the investigation of the transformation properties of the quadratic forms associated with (\ref{1h}) in \cite{VGG, Ner83}.
    Those objects turned out to be intimately related to the representation theory of the group $SL(2,R)$ and the Virasoro algebra.
    It would be natural to expect that such an invariance may play an important role in building a mathematically rigorous theory of path integrals of the type (\ref{1i}).}. Nevertheless, we eventually will be able to employ heuristic methods to conjecture the explicit value of  (\ref{1i})
    in a regularized versions of the theory for the particular functional $\Phi(V)=e^{-p\int_0^{2\pi}e^{-\beta V(t)}\,dt}$ with real parameters $p>0$ and $|\beta|<1$; see (\ref{1fpathint}) below.

  The similarity between (\ref{1/f}) and (\ref{8int}) is very significant in motivating the analogy we wish to draw between the statistical mechanical problem under discussion and the value distribution of the characteristic polynomials of CUE matrices, especially in the light of the discussion following (\ref{8int}). Unfortunately,  the random Fourier series (\ref{1/f}) is pointwise divergent with probability one, as reflected in
 the logarithmic divergence of the covariance (\ref{logcovarcont}) when $t_1\to t_2$. To be able to use such a function as a random energy landscape
 it is therefore necessary to provide a regularized version of the process with a well-defined variance\footnote{ Note that precisely the same process and the associated Boltzmann measure was found recently to play an important role in constructing
 the two-dimensional closed conformally-invariant random curves \cite{AJKS}}.
 To this end,  Fyodorov $\&$ Bouchaud \cite{FB08} constructed a lattice version of the model, called
the {\em circular-logarithmic} model.  In this case one replaces the function $V(t), \, t\in[0,2\pi)$ with a sequence of $M$ random mean-zero Gaussian variables $V_i$ obtained by sampling the 2dGFF equidistantly along the unit circle at points $z_j=\exp(i \frac{2 \pi j}{M})$ and then multiplying
by a constant factor $\sqrt{2\pi}$;  that is $V_i\equiv \sqrt{2\pi}\,V\left(\frac{2 \pi j}{M}\right), \,, i=1,\ldots, M$.
 Such a sequence automatically inherits the $M\times M$  covariance matrix $C_{km}=\mathbb{E}\{V_k V_m\}$ from (\ref{logcovarcont})
so that the off-diagonal entries are given by
\be \label{logcovardisc}
C_{k\ne m}=-2\log{|2\sin{\frac{2\pi}{M}(k-m)}|}\,.
\ee
To have a well-defined collection of Gaussian-distributed random variables we have to ensure the positive definiteness of the covariance matrix
by choosing the appropriate diagonal entries $C_{kk}$. A simple calculation shows that one has to choose
 \be \label{logvardisc}
 C_{kk}=\mathbb{E}\{V_k^2\}=2\log{M}+W_k, \quad \mbox{with any }\,\ \, W_k>0, \, \forall k=1,\ldots, M \,.
\ee
In practice it was further assumed that $W_k=W, \, \forall k$, and finally the limit $W\to 0$ was taken.

The crucial observation made in \cite{FB08}  was that  in the large-$M$ limit the positive integer moments $\mathbb{E}\{Z^n(\beta)\}$
of the partition function $Z(\beta)=\sum_i e^{-\beta V_i}$ corresponding to such a landscape converge
in a certain temperature range to the Dyson-Morris version of the Selberg Integral, see \cite{FW}.
More precisely, for $\beta<1$ and positive integer $n$ we have, asymptotically, that
\be\label{logcircmom}
\mathbb{E}\left\{ Z^n(\beta)\right\}=\left\{\begin{array}{cc}M^{1+\beta^2 n^2}\,O(1) & \quad \mbox{for}\,\, n>1/\beta^2\\
M^{n(1+\beta^2)} \frac{\Gamma(1-n\beta^2)}{\Gamma^n(1-\beta^2)}& \quad \mbox{for}\,\, 1<n<1/\beta^2\end{array}\right.
\ee
where $\Gamma(x)$ is the Euler Gamma-function.

From the moments
$ \mathbb{E}\{ Z^n(\beta)\}$ Fyodorov $\&$ Bouchaud were able to reconstruct the probability density ${\cal P}(Z)$ of the
partition function $Z({\beta})$ in the high-temperature phase
$0<\beta<1$. Actually all the statistical properties of the partition function remain invariant when changing $\beta\to -\beta$, so
we can formally consider $\beta$ to be of arbitrary sign, and define the high-temperature phase by the condition $|\beta|<1$. The density ${\cal P}(Z)$ in this domain turned out to consist of two pieces, the {\em body} and the {\em far tail}.  The  body of the distribution has a pronounced maximum at $Z \sim Z_e=M^{1+\beta^2}/\Gamma(1-\beta^2)\ll M^2$ and is given explicitly by
\be \label{partdiss}
 {\cal P}( Z)=\frac{1}{\beta^2}\frac{1}{Z}\left(\frac{Z_e}{Z}\right)^{\frac{1}{\beta^2}}\,
e^{-\left(\frac{Z_e}{Z}\right)^{\frac{1}{\beta^2}}},\quad  Z\ll M^2, \quad |\beta|<1
\ee
The most important feature of the above distribution is the powerlaw decay $ {\cal P}( Z)\sim Z^{-1-\frac{1}{\beta^2}}$ in the parametrically wide region $Z_e\ll  Z\ll M^2$. For $Z\gg  M^2$ the above expression is replaced by a lognormal tail
 \be \label{lognormal}
 {\cal P}(Z)=\frac{M}{\sqrt{4\pi \beta^2\log{M}}}
\frac{1}{Z}R\left(\frac{1}{2}
\frac{\log{Z}}{\log{M}}\right)\,e^{-\frac{1}{4\log{M}\beta^2}\ln^2{Z}}\,\quad \,,
\ee
where the unknown function $ R(x)$ is of the order of unity for $ x\sim O(1)$. It is easy to check that (\ref{partdiss}) and (\ref{lognormal})
indeed match at $Z\sim M^2$. Let us stress once again that this picture is valid only in the ``high-temperature" phase $|\beta|<\beta_c=1$.
%The sketch of this distribution is given in the figure.
% \begin{figure}[h!]
 %{\small }
%\begin{center}
%\includegraphics[width=7.0in]{yf.eps}
%\caption\label{Figure1}
%\end{center}
%\caption{Sketch of the probability density of the partition function in the high temperature phase. The parameter $\gamma=\beta^2$.}
%\end{figure}
%To understand
%the nature of what happens in the "low-temperature phase" $\beta>1$ we define $z={\cal Z}/{\cal Z}_e$ and consider the generating function
%\be \label{9}
%g_\beta(x) = \left\langle \exp( - e^{\beta x} z ) \right\rangle_{N\gg 1},
%\ee

\subsection{1/f-noise as a random landscape II: fluctuating multifractal patterns of heights and the threshold of extreme values}

In this section we review the arguments from \cite{FLeDR12}
allowing one to exploit the high-temperature moments (\ref{logcircmom}) to analyze the statistics of the number of ``low" or ``high" values in such a sequence. In this way we will be able to determine the position of the thresholds $V_{\pm}$ of the extreme values. For the high values, such a threshold $V_{+}$ is defined as the level above which typically we should have only a few (i.e.~of the order of one) points of the sequence $V_1,\ldots, V_M$ when $M\gg 1$. The definition of $V_{-}$ for low values is similar, with "above" replaced by "below". We already have mentioned that at the leading
order we must have $V_{\pm}=\pm 2\log{M}$. Correspondingly, we will call the value $V_i$ of the sequence $x-${\it high} provided $V_i=2x\log{M}$
with $0<x<1$, and similarly define $x-${\it low} values as those for which $-1<x<0$. Introducing further the notation $h_i=e^{V_i}>0$ the condition
  $V_i>x\ln{M}$ becomes equivalent to $h_i>M^{x}$. In the literature the general sets of values $h_i$ associated with points of the lattice in such a way that they scale in the large-$M$ limit as $h_i=M^{x_i}$, with {\it singularity exponents} $x_i$ filling in a finite interval $[x_{-},x_{+}]$ are called {\it multifractal} sets.  Their most important characteristic is the so-called {\it singularity spectrum} $f(x)$ which characterizes the asymptotic growth of the counting function $N_{>}(x)\sim M^{f(x)}$ of the number of $x$-high points. Thus, counting $x-$ high/low values is intimately related to revealing the multifractal structure of $1/f$ noise, and we give a brief account of the procedure below. This will help in quantifying the picture outlined in Section {\it 2.1}.

To that end we define the density $\rho_M(y)=\sum_{k=1}^M\delta(V_k-y\ln{M})$ in terms of which the counting function is given by $N_{>}(x)=\ln{M} \int_x^{\infty}\,\rho_M(y)\,dy$. In the large-$M$ limit the density can be described by the following {\it multifractal  Ansatz}:
\be\label{multiansatzdenhigh}
 \rho_M(y)\approx \frac{n_M(y)}{2\sqrt{\pi}\Gamma(1-y^2/4)} \frac{M^{f(y)}}{\sqrt{\ln{M}}},  \quad f(y)=1-\frac{y^2}{4},\,\,\, |y|<2
\ee
where $n_M(y)$ is a random coefficient of order of unity which  fluctuates strongly from one realization of the sequence $V_i$ to another.
To understand (\ref{multiansatzdenhigh}) we note that $\rho_M(y)$ provides a direct link between $N_{>}(x)$ and the partition function $Z(\beta)=\sum_i e^{-\beta V_i}$, as the latter can obviously be expressed in terms of the same function as
$Z(\beta)=\ln{M} \int_{-\infty}^{\infty}M^{-\beta y}\rho_M(y)\,dy\,$. Here it will be convenient to allow $\beta$ to be of any sign.
Substituting the Ansatz (\ref{multiansatzdenhigh}) into the above formula for $Z(\beta)$ we can perform the integral over $y$ in the limit $\ln{M}\gg 1$ by the Laplace method, with the stationarity condition resulting in the relation $y=-2\beta$. We arrive at the asymptotic relation $Z(\beta)\approx n_M(y=-2\beta) \, Z_e,$ with  $Z_e=\frac{M^{1+\beta^2}}{\Gamma(1-\beta^2)}$. Note that the condition $|y|<2$ translates into $|\beta|<1$. At the same time we know that  $Z(\beta)$ for $|\beta|<1$ must be distributed according to the density (\ref{partdiss}). To ensure this property we therefore conclude that the probability density of the random variable $n=n_M(y)$ for a fixed value of $y\in(-2,0)\bigcup (0,2)$  must necessarily be of the form
\be \label{dissnumb}
 {\cal P}_y(n)=\frac{4}{y^2}\frac{1}{n^{1+\frac{4}{y^2}}}\,
e^{-\left(\frac{1}{n}\right)^{\frac{4}{y^2}}},\quad  y\in(-2,2),\,\, y\ne 0.
\ee
We expect this form of the density to be valid as long as $n\ll n_c$, with $ n_c$ being some cutoff value diverging for $M\to \infty$.
The precise dependence of $n_c$ on $M$, as well as the shape of ${\cal P}_y(n)$ for $n\gg n_c$, cannot be extracted from the above arguments,
although the internal consistency of the Laplace method suggests that (\ref{dissnumb}) cannot hold when $n\sim M^{\nu}$ when $\nu>0$.

  These facts can now be used to determine the counting function statistics. Indeed, substituting (\ref{multiansatzdenhigh}) into the integral for the counting function we find by the same method $N_{>}(x)\approx  n_M(x)\, {\cal N}_t(x)\,$ with the same random factor $n_M(x)$ distributed according to (\ref{dissnumb}) and the typical value ${\cal N}_{t}(x)$ being given by
\be\label{charscale}
{\cal N}_{t}(x)= \frac{M^{1-x^2/4}}{x\sqrt{\pi\ln{M}}}\frac{1}{\Gamma(1-x^2/4)}\equiv \mathbb{E}\left\{N_{>}(x)\right\}\frac{1}{\Gamma(1-x^2/4)}, \quad 0<x<2\,.
\ee
The significance of the above is, in particluar, the fact that it allows one to determine the precise position
 $V_{+}$ of the typical threshold of extreme values. By definition the threshold is determined by the condition ${\cal N}_{t}(x)\sim 1$
 when $M\gg 1$.  A straightforward calculation then allows one to show that
\be \label{threshold}
V_{+}=2x_{+}\ln{M}, \quad \mbox{where} \quad x_{+}=2-c\,\frac{\ln{\ln{M}}}{\ln{M}}, \quad \mbox{with} \quad c=3/2
\ee
The same large-$M$ asymptotic must also hold for the position of the absolute maximum $V_m$ of the sequence which is always among a few values above that threshold. The statistical properties of $V_m$ will be discussed in more detail in the next section.

Note finally that had we instead decided to use the condition
 $\mathbb{E}\left\{N_{>}(x)\right\}\sim 1$ this would result again in (\ref{threshold}) but with $c=3/2$ replaced by $c=1/2$.
 The latter value is indeed known to be characteristic of short-ranged correlated random sequences.
The difference is due to the fact that for such sequences the mean and the typical values of the counting function
are always of the same order, whereas for log-correlated sequences in the vicinity of the threshold
 the typical value ${\cal N}_{t}(x)$  becomes parametrically smaller than the mean value $\mathbb{E}\left\{N_{>}(x)\right\}$.
 Similar behaviour is believed to be shared by a broad class of disorder-dominated multifractal processes and fields, see \cite{FLeDR12} for a more detailed discussion.

\subsection{1/f-noise as a random landscape III: duality, freezing and statistics of extremes}

 Now we turn our attention to the generating function $ \mathbb{E}\left\{ \exp( - e^{\beta y} Z/Z_e  ) \right\}$ which will underpin many subsequent calculations. It may be checked easily that the leading-order large-$M$ behaviour is dominated by the ``body" density (\ref{partdiss})
 rather than by the log-normal tail (\ref{lognormal}), and so we find after straightforward manipulations:
\begin{eqnarray}\label{mainlogcirc}
&&  g_\beta(y)=\mathbb{E}\left\{ \exp( - e^{\beta y} Z/Z_e  ) \right\}\equiv \mathbb{E}\left\{ \exp( - e^{\beta \left(y-\phi_{\beta}\right) }\right\}
\\&& = \int_0^{\infty} \exp\left\{-t-e^{\beta y}t^{-\beta^2}\right\}dt, \quad 0<\beta<1 \nonumber
\end{eqnarray}
where we have employed the notation $\phi_{\beta}=F(\beta)-F_e(\beta)$ for the deviation of the free energy $F=-\beta^{-1}\ln{Z(\beta)}$ from its typical value in the high-temperature phase
\be\label{fe}
F_e=-\beta^{-1}\ln{Z_e}=-(\beta+\beta^{-1})\ln{M}-\beta^{-1}\ln{\Gamma(1-\beta^2)}
\ee

Note that after identifying $M^{-1}Z(\beta)$ as a regularization for the $1/f$ noise integral $\frac{1}{2\pi}\int_0^{2\pi}e^{-\beta V(t)}\,dt$,
(\ref{mainlogcirc}) can be interpreted as the evaluation of the ``path integral" (\ref{1h})-(\ref{1i}) for the functional $\Phi(V)=e^{-p\frac{1}{2\pi}\int_0^{2\pi}e^{-\beta V(t)}\,dt}$ with $p>0$, and in this way is equivalent to the identity:
\be\label{1fpathint}
 \mathbb{E}\left\{e^{-p\frac{1}{2\pi}\int_0^{2\pi}e^{-\beta V(t)}\,dt}\right\}=\int_0^{\infty}\,e^{-t-\frac{p z_e(\beta^2)}{t^{\beta^2}}}dt, \quad z_e(\beta^2)=\frac{e^{\frac{\beta^2}{2}{\small \mathbb{E}\{V^2(t)\}}}}{\Gamma(1-\beta^2)}
\ee
expected to hold when $|\beta|<1$ for any regularized version of the $2\pi$-periodic Gaussian $1/f$ noise with finite variance $\mathbb{E}\{V^2(t)\}<\infty$.
In particular, let us take the limit $p\to \infty, \beta \to 0$ in such a way that $p\beta^2=\mu<\infty$.
Note that for the periodic $1/f$ noise (\ref{1/f}) we obviously must have $\int_0^{2\pi} V(t)\,dt\equiv 0$. Using this fact we easily find that, in the limit in question, (\ref{1fpathint}) yields the identity:
\be \label{1frough}
  \mathbb{E}\left\{e^{-\frac{\mu}{2\pi}\int_0^{2\pi} V^2(t)\,dt}\right\}=e^{-\mu z'_e(0)} \Gamma(1+\mu)
\ee
where $z'_e(0)=\frac{d}{d(\beta^2)}z_e(\beta^2)|_{\beta^2=0}=\frac{1}{2} \mathbb{E}\{V^2(t)\}+\Gamma'(1)$ is a constant depending on the chosen regularization. In turn, (\ref{1frough}) means that the quantity $R=\frac{1}{2\pi}\int_0^{2\pi} V^2(t)\,dt$ (which can be interpreted as
a measure of {\it roughness} of the $1/f$ signal) is Gumbel-distributed: $R= z'_e(0)+r$ with random $r$ whose probability density is ${\cal P}(r)=\frac{d}{dr}\exp\left\{-e^{-r}\right\}$. This result was derived for the first time in \cite{ADGR} by a completely different method, and remained, until recently, one of only a few explicit results on the statistics of $1/f$ noise.

The integral on the right-hand side of (\ref{mainlogcirc}) belongs to a class of special functions that enjoyed a detailed investigation
in the mathematical literature only a few years ago, see \cite{Ner06}. It possesses several non-trivial properties.
 In particular, it was noticed in \cite{FLeDR09} that $g_\beta(y)$  satisfies a remarkable and important {\em duality relation}:
$g_{\mathbf{\beta}}(y)=g_{\frac{1}{\beta}}(y)\,$. Indeed, after some algebraic manipulations one can rewrite the integral in the right-hand side
 as
\begin{eqnarray}\label{mainlogcirc1}
 \int_0^{\infty}  \exp\left\{-t-e^{\beta y}t^{-\beta^2}\right\}dt =\frac{1}{2\pi i}\int_{Re v=\epsilon>0}e^{-vy}\Gamma(1+\beta v)\Gamma\left(1+\beta^{-1} v\right)\,\frac{dv}{v}\\
=1+\sum_{n=1}^{\infty}\frac{(-1)^n}{n!}\left[e^{n\beta y}\Gamma(1-n\beta^2)+e^{n\frac{y}{\beta}}\Gamma\left(1-\frac{n}{\beta^2}\right) \right],
\end{eqnarray}
where in the right-hand side formally $ 1<\frac{1}{\beta^{2}}\ne \mbox{integer}$, but it is easy to check that in the limit
 $\frac{1}{\beta^{2}}\to \mbox{integer}$ the series in fact retains a well defined finite value.

Let us again stress that (\ref{mainlogcirc})  holds only for $0<\beta<1$ and the duality of the right-hand side cannot be used to
  evaluate the left-hand side for $\beta>\beta_c=1$. Instead, there is accumulating evidence that a type of {\it freezing} phase transition happens at $\beta=\beta_c$. We have already mentioned the simplest instance of that transition at the level of the mean free energy which manifests itself (to leading order in $M$) in the change from the value $-(\beta+\beta^{-1})\ln{M}$ for $\beta<1$ to the temperature-independent
  value $-2\ln{M}$ for $\beta>1$. This freezing of the leading-order mean free energy, which follows the pattern of Derrida's uncorrelated REM, was, for the present model, rigorously proved recently in \cite{AZ}. However,  the idea of freezing has been elevated far beyond the level of the first moment and conjectured to extend to
  a much richer {\em freezing scenario}: the whole generating function $ g_\beta(y)$ 'freezes' to the temperature independent profile $ g_{\beta=1}(y)$ everywhere in the 'glassy' phase $\beta>1$ \cite{CLeD, FB08}.

This scenario is supported by the following arguments
 (i) heuristic real-space renormalization group calculations \cite{CLeD}  revealing an analogy to the
travelling wave analysis of polymers on disordered trees \cite{DS} where such a scenario can be rigorously shown to
hold (ii) the duality relation mentioned above can be shown to hold also for other types of logarithmic landscapes \cite{FLeDR09, FLeDR10} -- in particular, the duality forces the 'temperature flow' of the function $ g_\beta(x)$
to stop at the critical point $\beta=\beta_c=1$ -- (iii) by relations between the freezing scenario and the mechanism of one-step replica symmetry breaking in logarithmic models \cite{FLeDR10}, and (iv) finally, by direct numerical simulations in these papers. All these facts taken together inspire our confidence in the validity of the freezing scenario,
although it remains a major challenge to prove this conjecture rigorously.

One of the main consequences of the freezing conjecture  is that it implies the possibility of obtaining the distribution
of the (properly rescaled) free energy in the low-temperature phase. Namely, following \cite{FB08} we
make an additional assumption that for all $\beta>1$ when freezing is operative the expression (\ref{fe}) for the typical value of the free energy $F_{e}(\beta)$ should be replaced with the value of the position of the threshold of minimal values $F_{e}(\beta)\to V_{-}=-2\ln{M}+\frac{3}{2}\ln{\ln{M}}$, so that $\phi_{\beta}=F(\beta)-V_{-}$. Introducing again the notation $p=e^{\beta y}$ and using the fact that
for our particular choice of the model $g_{\beta=1}(y)=2 e^{y/2}K_1(2 e^{y/2})$ where $K_1(z)$ denotes the modified Bessel function of the second kind, we see that the freezing scenario allows us to rewrite the relation (\ref{mainlogcirc}) for all $\beta>1$ as
\be\label{freez}
 \mathbb{E}\left\{\exp{\left(-pe^{\phi_{\beta}}\right)}\right\}=2p^{\frac{1}{2\beta}}K_1\left(2p^{\frac{1}{2\beta}}\right),
\ee
This formula can be used as a generating function for the random variable $\phi=\phi_{\beta}$, and can be further employed
to extract the probability density for that variable in a closed form:
\be\label{12} {\cal P}_{\beta>\beta_c}^{CLM}(\phi)=
\frac{1}{2\pi}\int_{-\infty}^{\infty}\,e^{-is \phi}\,
\frac{1}{\Gamma(1+\frac{is}{\beta})}
\Gamma^2\left(1+is\right)\,ds
\ee
$$
=-\frac{d}{d\phi}\left[1+\sum_{n=1}^{\infty}\frac{e^{n\phi}}{n!(n-1)!\Gamma\left(1-n\frac{1}{\beta}\right)}
\left(\phi+\frac{1}{n}-2\psi(n+1)+\frac{1}{\beta}\psi\left(1-n\frac{1}{\beta}\right)\right)\right]\,,
$$
 where $\psi(x)=\Gamma'(x)/\Gamma(x)$.
In particular, as $\lim_{\beta\to \infty}\left[-\beta^{-1}\log{Z(\beta)}\right]=\mbox{min}_{i=1,\ldots,M}\{V_i\}=V_{min}$, one can gain access to the
distribution of the minimum of the random potential sequence. Introducing, correspondingly, $x=\lim_{\beta\to \infty}\phi(\beta)=V_{min}+2\log{M}-\frac{3}{2}\ln{\ln{M}}$ one finds in the large-$M$ limit the  probability density for the variable $x$ to be
\begin{eqnarray}  \label{circ}
p(x)=-\frac{d}{dx}\left[2 e^{x/2}
K_1(2 e^{x/2})\right]=2 e^{x}
K_0(2 e^{x/2})\,.
\end{eqnarray}
 Such a distribution is manifestly different from the Gumbel law, which has a probability density of the form
 $p_{Gum}(x)=-\frac{d}{dx}\left[\exp{-Be^{Ay}}\right]$ and which holds for short-range correlated gaussian random sequences.
Moreover, the density $p(x)$ does indeed exhibit the universal Carpentier-Le Doussal tail $p(x\to -\infty) \sim - x e^x$.
One can easily find all cumulants of the distribution (\ref{circ}), see \cite{FLeDR09}, e.g. the mean $\mathbb{E}\{x\}=-2\gamma_E$, the variance $\left[\mathbb{E}\{x^2\}\right]_c=\mathbb{E}\{x^2\}-\mathbb{E}\{x\}^2=\frac{\pi^2}{3}$, etc., $\left[\mathbb{E}\{x^n\}\right]_c=(-1)^n 2(n-1)!\zeta(n)$ in terms of Riemann-zeta values. All this agrees with the available numerics.

\section{Statistics of extreme and high values of CUE characteristic polynomials.}

The apparent similarity between the circular-logarithmic model discussed in the previous section, and the logarithm of the characteristic polynomial of a CUE matrix makes it evident that they exemplify two different ways of regularizing the same $2\pi-$periodic $1/f$-noise (\ref{1/f}).
The latter can be properly defined only as a random generalized function,  see \cite{Sheff}.
The two regularizations are however of a rather different nature: the log-circular model replaces   $1/f$ noise
 with a finite sequence of $M$ random variables, whereas the log-modulus of the characteristic polynomial for any finite $N$ is a  piecewise-continuous random function with $N$ logarithmic singularities in the interval $[0,2\pi]$. Nevertheless, if the limiting $1/f$ noise is a meaningful object at all, it would be natural to expect the two processes to share the same extremal and high/low-level values statistics in the limits $\log{M}\gg 1$ and $\log{N}\gg 1$, respectively.

To substantiate this claim we follow the same strategy for characteristic polynomials as we did for the log-circular models.
Our main object of interest in this section will therefore be the moment (\ref{1}), which is analogous to the partition function discussed in the previous section, and in the associated free energy.
We shall be interested in various choices of the interval $L$.

The first step is to consider the positive integer moments
\be\label{3}
\fl \mathbb{E}\left\{{\cal Z}_N^k(\beta;L) \right\}=N^k\int_0^{L}\ldots \int_0^{L}  \mathbb{E}\left\{ |p_N(\theta_1)|^{2\beta}\ldots |p_N(\theta_k)|^{2\beta}\right\}\prod_{j=1}^k\frac{d\theta_j}{2\pi}, \quad k=1,2,\ldots
\ee
where the expectation $\mathbb{E}\{\ldots\}$ is with respect to Haar measure on the unitary group ${\cal U}(N)$. As is well known, the expectation value in the integrand is given by a Toeplitz determinant:
\be\label{4}
 \mathbb{E}\left\{ |p_N(\theta_1)|^{2\beta}\ldots |p_N(\theta_k)|^{2\beta}\right\}=\frac{D_N^{(k)}(\beta)}{D_N^{(k)}(0)}, \quad
 D_N^{(k)}(\beta)=\det{\left(M_{i-j}^{(\beta)}\right)_{i,j=0}^{N-1}}
\ee
where
\be\label{5}
M_{i-j}^{(\beta)}=\int_0^{2\pi} e^{i\phi (i-j)}\prod_{p=1}^{k}\left[2-2\cos{(\phi-\theta_p)}\right]^{\beta}\frac{d\phi}{2\pi}.
\ee
 The asymptotic behaviour of such  a Toeplitz determinant in the limit $N\to \infty$  was conjectured in \cite{FH} and proved in \cite{Wid}:
 \be\label{6}
D_N^{(k)}\approx \left[N^{\beta^2}\frac{G^2(1+\beta)}{G(1+2\beta)}\right]^k \prod_{r<s}^k|e^{i\theta_r}-e^{i\theta_s}|^{-2\beta^2}
\ee
where $G(x)$ is the Barnes function. Further progress is possible in two cases.

\subsection{The full circle: $L=2\pi$}

In this case, substituting (\ref{6}) back to  (\ref{4}) and (\ref{3}), we see that the resulting expression is the standard Dyson-Morris
version of the Selberg integral, see \cite{FW},
 convergent for $k<\frac{1}{\beta^2}$ and divergent for larger $k$. As we have $k\ge 1$ the procedure makes sense only for $\beta^2<1$.
 Introducing the notation ${\cal Z}_e=N^{1+\beta^2}\frac{G^2(1+\beta)}{G(1+2\beta)\Gamma(1-\beta^2)}$, we find
 \be\label{7}
\mathbb{E}\left\{{\cal Z}_N^k(\beta) \right\}= {\cal Z}_e^k \Gamma(1-k\beta^2), \quad k<\frac{1}{\beta^2}.
\ee

 To understand how to deal with the case $k>\frac{1}{\beta^2}$, we recall that Widom's asymptotic formula (\ref{6})
 is valid only when all of the differences $|\theta_i-\theta_j|$ remain {\it finite} when $N\to \infty$, and should be replaced by a different
 expression when  $ |\theta_i-\theta_j|\sim N^{-1}$. One can check that the divergence of the integral for $k>1/\beta^2$ is due precisely to the fact
 that these near degeneracies become important. Relying on our experience with the corresponding situation for the circular-logarithmic case
  suggests that taking into account the correct short-scale cutoff  cures the formal divergence, but changes the asymptotics of the moments $\left\langle {\cal Z}_N^k(\beta) \right \rangle$ with $N$: namely, these become of the order of $N^{1+k^2\beta^2}$ for
 $k>\beta^{-2}$ whereas they are of the order of $N^{(1+\beta^2)k}$ for $k<\beta^{-2}$ (c.f.~the large-$N$ asymptotics of the moments of the characteristic polynomials derived in \cite{KS}). Such a change of behaviour will lead to a log-normal (far) tail in the distribution.

 The above expression (\ref{7}) for the moments is exactly the same as (\ref{logcircmom}), except that
 the characteristic scale ${\cal Z}_e$ has a different value, with the ratio of the two scales given by the factor $\frac{G^2(1+\beta)}{G(1+2\beta)}$. Note that the factor tends to unity when $\beta$ approaches the freezing temperature $\beta=1$. Relying then on the freezing paradigm we conclude that such a factor will not affect the statistics of the maximum, and we can simply translate, {\it mutatis mutandis}, all results from
 the regularized circular-logarithmic model to the values of characteristic polynomial sampled along the full circle $\theta\in(0,2\pi]$.

\subsection{A mesoscopic arc: $1\ll N_L=N\frac{L}{2\pi}\ll N$}

Another case when the partition function moments can be explicitly evaluated corresponds to an arc of
length $L$ that is much smaller than that of the full circle, but still much larger than the typical distance between the eigenvalues,
$1/N$.  We will call such an intermediate scale {\em mesoscopic}.
Following the same route as before, in the limit $N\gg 1$ we have
\be\label{15}
\fl \mathbb{E}\left\{{\cal Z}_N^k(\beta,L) \right \}\approx \left[N^{1+\beta^2}\frac{G^2(1+\beta)}{G(1+2\beta)}\right]^k
\int_0^{L}\ldots \int_0^{L}  \prod_{r<s}^k|e^{i\theta_p}-e^{i\theta_s}|^{-2\beta^2}\prod_j \frac{d\theta_j}{2\pi}
\ee
which after expanding the exponents $e^{i\theta_p}\approx 1+i\theta_p$, and further rescaling
$\theta_p=L y_p$ acquires the form
\be\label{16}
\fl \mathbb{E}\left\{ {\cal Z}_N^k(\beta) \right \}\approx \left[\frac{N^{1+\beta^2}}{2\pi}\frac{G^2(1+\beta)}{G(1+2\beta)}\right]^k L^{-\beta^2k(k-1)+k}
\int_0^{1}\ldots \int_0^{1}  \prod_{r<s}^k|y_p-y_s|^{-2\beta^2}\prod_j\,dy_j
\ee
Notice that the factor $L^{-\beta^2k^2}$ can be written as $\mathbb{E}\left\{\left[e^{\frac{1}{2}u\sqrt{-2\ln{L}}}\right]^{2\beta k}\right\}$,
where the averaging is performed over the standard mean-zero gaussian variable  $u$ of unit variance.  We conclude that the  characteristic polynomial $p_{N}(\theta)$
in the interval $[0,L]$ has the same probability law as a product of two independent random variables:  $p_{N}(\theta)=e^{\frac{1}{2}u\sqrt{-2\ln{L}}} \times\tilde{p}_{N}(\theta)$, such that for the 'reduced' partition function $\tilde{{\cal Z}}_{N}(\beta;L)=\frac{N}{2\pi}\int_0^{L}|\tilde{p}_N(\theta)|^{2\beta}d\theta$ we have
\be\label{16a}
\mathbb{E}\left\{ \tilde{{\cal Z}}_N^k(\beta) \right \}\approx \left[\frac{(NL)^{1+\beta^2}}{2\pi}\frac{G^2(1+\beta)}{G(1+2\beta)}\right]^k
\int_0^{1}\ldots \int_0^{1}  \prod_{r<s}^k|y_p-y_s|^{-2\beta^2}\prod_j\,dy_j
\ee
The integral in the above equation is the standard Selberg integral \cite{FW}, which gives finally
\be\label{17}
 \mathbb{E}\left\{ \tilde{{\cal Z}}_N^k(\beta) \right \}\approx \left[\tilde{{\cal Z}}_e\right]^k\prod_{j=1}^{j=k} \frac{\Gamma^2[1-(j-1)
\beta^2] \Gamma(1-j
\beta^2)}{\Gamma[2-(k+j-2) \beta^2]},\quad 1<k<\beta^{-2},
\ee
where we have introduced the scale  $\tilde{{\cal Z}}_e= N_L^{1+\beta^2}(2\pi)^{\beta^2}\frac{G^2(1+\beta)}{G(1+2\beta)\Gamma(1-\beta^2)}$,
with $N_L\equiv N\frac{L}{2\pi}$.
For the case $k>\beta^{-2}$ the integral is divergent and we can apply a consideration similar to that applied for the full-circle case, which
 is of secondary importance for our goals and will not be repeated here.

We see that the expressions (\ref{17}) for the moments of  $z(\beta)=\tilde{{\cal Z}}_N/\tilde{{\cal Z}}_e$
are  precisely the moments of the distribution analyzed in \cite{FLeDR09}.
That paper dealt with the aperiodic version of the $1/f$-noise sampled from the two-dimensional full-plane GFF along an interval
of unit length. The analysis followed essentially the same steps as for the circular-logarithmic model, but technically was much more involved due to the complicated structure of the moments. The main difficulty was to find a method of continuing $ \mathbb{E}\left\{ \left[{z(\beta)}\right]^k \right \}$ to complex
$k$ in a meaningful way, which allows one to find the probability density of $z$ from the above moments. The goal was successfully
achieved in  \cite{FLeDR09} by employing heuristic methods.
 Independently, a very similar problem arose in the context of financial mathematics, and an elegant mathematically
rigorous solution was proposed in \cite{Ost09, Ost12}, with the results of the two approaches coinciding.

The extreme/high value statistics for the modulus of the characteristic polynomial in the interval $N^{-1}\ll \frac{L}{2\pi}\ll 1$
can again be translated straightforwardly from those results, and was summarized in the beginning of Section 2.

\subsection{Multifractal properties of the modulus of characteristic polynomials}

 The results for  mesoscopic intervals presented above can be also interpreted
  as revealing the multifractal structure of the function $|p_N(\theta)|$.
 To see this more clearly we resort to a standard tool of multifractal analysis, {\it box counting}, and
 subdivide  the full interval $[0,2\pi]$ into $M=2\pi/l_b$ subintervals ("boxes") of equal length $l_b$ chosen in such a way that
 the number of boxes satisfies $1\ll M=2\pi/l_b\ll N$. We further define the set of coarse-grained values $h_n(l_b), n=1,\ldots,M$ of the function $|p_N(\theta)|$ by averaging it over the $n-$th box and the corresponding "partition functions" $\zeta_q(l_b)$ via
 \be\label{av1}
 h_n(l_b)=\frac{1}{l_b}\int_{(n-1)l_b}^{nl_b} |p_N(\theta)|\,d\theta, \quad \zeta_q(l_b)=\frac{1}{M}\sum_{i=1}^{M}
 \left[h_n(l_b)\right]^q
\ee
Now, straightforward calculation gives for integer $q$ the ensemble-averaged value:
\be\label{av2}
\fl \mathbb{E}\left\{\zeta_q(l_b\right\}=M^q\int_{0}^{l_b}\mathbb{E}\left\{ |p_N(\theta_1)|\ldots
|p_N(\theta_q)|\right\}\,\frac{d\theta_1}{2\pi}\ldots \frac{d\theta_q}{2\pi}=\left( \frac{M}{N}\right)^q\mathbb{E}\left\{{\cal Z}_N^q(\beta=1/2;l_b) \right\}
\ee
\be\label{av3}
=\left[\frac{N^{1/4}G^2(3/2)}{G(2)}\right]^q S_q(\beta^2=1/4)\, l_b^{-\frac{1}{4}q(q-1)}, 1\le q<4
\ee
where we have used the rotational invariance of the ensemble-averaged values, the definition (\ref{3}) and the formula (\ref{16}),
with $S_q(\beta^2)$ standing for the $\beta$-dependent Selberg integral featuring in (\ref{16}) and convergent for $1<q<\beta^{-2}$.
Though formally derived for the integer $q$ one expects the above to be valid for real $q<4$ as well, so we finally
arrive to the relation
 \be\label{av4}
\fl i_q(l_b)=\frac{\mathbb{E}\left\{\zeta_q(l_b\right\}}{\left[\mathbb{E}\left\{\zeta_1(l_b\right\}\right]^q}=S_q(1/4)\,
l_b^{-\tau_q}, \quad \tau_q=\frac{1}{4}q(q-1),\quad\frac{1}{N}\ll l_b \ll 2\pi, \quad q<4
\ee
 In view of the nonlinear dependence of the exponent $\tau_q$ on $q$, this scaling of the ratio $i_q(l_b)$ associated with coarse-graining  on the scale $l_b$ implies multifractality of the underlying function $|p_N(\theta)|$, see e.g.~\cite{FLeDR12}
with discussion and further references therein.  In fact, multifractality could be immediately anticipated from the presence of the lognormal factor $e^{\frac{1}{2}u\sqrt{-2\ln{l_b}}}$ in the  ``mesoscopic" statistics of $|p(\theta)|$ with an exponent whose variance, $-\ln{l_b}\approx \ln{M}$, is linear in the log of the number of boxes.

\subsection{Measure of high points of characteristic polynomials}
The multifractal structure of $1/f$ noise also shows up in a somewhat different, but related way in the statistics of
high values of $|p_N(\theta)|$. Note that as follows from our discussion the typical value of the maximum of $|p_N(\theta)|$ in the interval $\theta\in[0,L]$ is $N_L$. The simplest quantity which helps to quantify the structure of high values is
the relative length  $\mu_N(x;L)$ (as fraction of the total length $L$) of those intervals in $[0,L]$ where $|p_N(\theta)|>(N_L)^x $ for a fixed $ 0<x<1$.
Statistics of this quantity are naturally related to the statistics of the partition function in the high-temperature phase $\beta<1$,
as was informally discussed for the case of the log-circular model in Section 3. To substantiate the claim for the characteristic polynomials
we choose here to follow an alternative procedure. Though some parts of the method remain of somewhat heuristic nature, we believe it grasps the mathematical structures correctly. It remains a challenge to
justify it by rigorous methods.

 We start from the definition
\be \label{m1a}
\mu_N(x;L)=\frac{1}{L}\int_0^{L}\chi\{2\log{|p_N(\theta)|-2x\log{N_L}\}}d\theta,
\ee
where $\chi\{u\}=1$ if $u>0$ and zero otherwise. Using the Fourier-transform one can rewrite this as
\be \label{m2}
\mu_N(x;L)=\frac{2\log{N_L}}{2\pi}\int_0^{L}\frac{d\theta}{L}\int_x^{\infty}\,d \eta\int_{-\infty}^{\infty}\, dk\,e^{-2ik\eta\log{N_L}}
 |p_N(\theta)|^{2ik}
\ee
  Our goal is again to calculate the integer moments $\left\langle\mu_N(x;L)^p\right\rangle $ for $p=1,2,\ldots$. These can be found by
  the same procedure we used for the ``partition function" moments above and are
   given by
  \be \label{m3}
\left\langle\mu_N(x;L)^p\right\rangle=\left(\frac{\log{N_L}}{\pi}\right)^p\int_0^{L}\frac{d\theta_1}{L}\ldots \int_0^{L}\frac{d\theta_p}{L}
\int_x^{\infty}\,d \eta_1\ldots d\eta_p \, {\cal I}(\eta_1,\ldots,\eta_p),
\ee
where
\be\label{m4}
\fl {\cal I}(\eta_1,\ldots,\eta_p)=\int_{-\infty}^{\infty}\,dk_1\ldots dk_p e^{-2i\log{N_L}\sum_{j=1}^p k_j\eta_j} \left\langle |p_N(\theta_1)|^{2ik_1}\ldots |p_N(\theta_p)|^{2ik_p}\right\rangle
\ee

The ensemble average is again a Toeplitz determinant:
\be\label{m5}
 \left\langle |p_N(\theta_1)|^{2\lambda_1}\ldots |p_N(\theta_k)|^{2\lambda_p}\right\rangle=\frac{D_N^{(p)}(\beta)}{D_N^{(p)}(0)}, \quad
 D_N^{(k)}(\beta)=\det{\left(M_{i-j}^{(\beta)}\right)_{i,j=0}^{N-1}}
\ee
where
\be\label{m6}
M_{i-j}^{(\beta)}=\int_0^{2\pi} e^{i\phi (i-j)}\prod_{j=1}^{p}\left[2-2\cos{(\phi-\theta_j)}\right]^{\lambda_j}\frac{d\phi}{2\pi}
\ee
As already noted, the asymptotic behaviour of such  a Toeplitz determinant in the limit $N\to \infty$  is given by the Fisher-Hartwig conjecture \cite{FH, Wid}:
 \be\label{m7}
D_N^{(p)}\approx \prod_{j=1}^p\left[N^{\lambda_j^2}\frac{G^2(1+\lambda_j)}{G(1+2\lambda_j)}\right] \prod_{r<s}^p|e^{i\theta_r}-e^{i\theta_s}|^{-2\lambda_r\lambda_s}
\ee
where $G(x)$ is the Barnes function, and the formula is valid for $\mbox{Re}\lambda_j>-1/2$. For purely imaginary $\lambda_j=ik_j$,
 which is our case, it gives
\be\label{m8}
\fl
{\cal I}(\eta_1,\ldots,\eta_p)=\int_{-\infty}^{\infty}\,dk_1\ldots dk_p e^{-\log{N_L}\sum_{j=1}^p (k_j^2 +2ik_j\eta_j)}
\prod_{j=1}^p\left[\frac{G^2(1+ik_j)}{G(1+2ik_j)}\right]\prod_{r<s}^p|e^{i\theta_r}-e^{i\theta_s}|^{2k_rk_s}
\ee
In the limit $\log{N_L}\gg 1$ the integral is dominated by the saddle point $k_j=-i\eta_j, \, \forall j=1,\ldots,p$ which gives
\be\label{m9}
\fl
{\cal I}(\eta_1,\ldots,\eta_p)\approx \left(\frac{\pi}{\log{N_L}}\right)^{p/2} e^{-\log{N_L}\sum_{j=1}^p \eta_j^2}
\prod_{j=1}^p\left[\frac{G^2(1+\eta_j)}{G(1+2\eta_j)}\right]\prod_{r<s}^p|e^{i\theta_r}-e^{i\theta_s}|^{-2\eta_r\eta_s}.
\ee
Substituting this back into (\ref{m3}) one may expect that all
$\eta-$integrals for $\log{N_L}\gg 1$ and fixed $x>0$ will be dominated by
the lower limit $\eta_j=x, \forall j$. To this end we introduce new variables $u_j$ by
$\eta_j=x+\frac{u_j}{2\log{N_L}}, \, \forall j=1,\ldots,p$ and find for $x\gg 1/\sqrt{\log{N_L}}$
\be \label{m10}
\fl
\left\langle\mu_N(x;L)^p\right\rangle=\left(\frac{\log{N_L}}{\pi}\right)^{p/2}\frac{N_L^{-px^2}}{(2x\log{N_L})^p}
\left[\frac{G^2(1+x)}{G(1+2x)}\right]^p
\int_0^{L}\frac{d\theta_1}{L}\ldots \frac{d\theta_p}{L}\,\prod_{r<s}^p|e^{i\theta_r}-e^{i\theta_s}|^{-2x^2}
\ee

The remaining integral can be performed explicitly again in the two limits:\\ (i) the full-circle case $L=2\pi$ so that $N_L=N$ and (ii)
the mesoscopic  interval $1\ll N_L\ll N$. In case (i) it is the familiar Dyson-Morris integral which gives $\Gamma(1-px^2)/[\Gamma(1-x^2)]^p$ for $0<x^2<1/p$ and
diverges otherwise (the same remark on the divergence of the moments of the partition function applies here). Collecting all factors, we get
\be \label{m11}
\left\langle\mu_N(x)^p\right\rangle|_{\log{N}\gg 1}\approx \left[\mu_e(x)\right]^p \Gamma(1-px^2), \quad 0<x^2<1/p
\ee
where the 'typical' value $\mu_e(x)$ for the measure is given by
\be \label{m12}
 \mu_e(x)=N^{-x^2}\sqrt{\frac{1}{\pi\log{N}}}\frac{G^2(1+x)}{2x \, G(1+2x)}\frac{1}{\Gamma(1-x^2)}, \quad 0<x<1
\ee
The nontrivial leading scaling with $N^{-x^2}$ reflects the multifractal-type structure of the measure of intervals supporting high values.

Note that the mean value of the measure is given by $\mathbb{E}\left\{\mu_N(x)\right\}=\mu_e(x)\Gamma(1-x^2)$.
 It obviously stays finite for $x \to 1$, and a direct calculation shows such an expression is valid for any
$x>0$, without restricting to $x<1$. However, when approaching $x=1$  the mean value
is significantly larger than the typical value, and is dominated by rare fluctuations.
Recall that $x_{+}=1-c\ln{\ln{N}}$ with $c=\frac{3}{4}$ is conjectured to be the thereshold of extreme values.
This claim is consistent with the fact that at such a level $x$ the measure supporting high values is of the order of $\mu_e(x)\sim N^{-1}$,
that is comparable with the minimal scale of the problem (the typical separation between zeroes) which simply means there are typically of order
of one maxima above such a level.

In the whole interval $0<x<1$ the probability density of $\mu\equiv\mu_N(x)$ can be immediately recovered
by noticing that the moments of $\xi=\mu/\mu_e(x)$ coincide with those of the random variable $Z/Z_e$ distributed according to (\ref{partdiss}),
with the obvious identification $x^2\to \beta^2$.
 We conclude that the total relative length $\mu_{N\gg 1}(x)$ of the intervals supporting high values of characteristic polynomial
is distributed according to the probability density ${\cal P}({\cal \mu})$ given by  (\ref{m13sum}).
Such an expression should be valid for all $\mu$ as long as $\mu \ll 1$ and when $\mu\sim 1$ must have a sharp cut off, as
obviously the fraction of the total length cannot be larger than unity in any realization. Note, however, that although the above picture is in good qualitative agreement with numerical experiments performed for the circular-logarithmic model,  it has so far not proved possible to achieve quantitative agreement with (\ref{m13sum})  for any realistic numerical
simulation of $1/f$ noise (see \cite{FLeDR12}).

For the mesoscopic interval with $L\ll 1$ we can proceed similarly, and, again
after expanding in (\ref{m10}) the integrand assuming $\theta_j\ll 1$,
 connect to the corresponding moments of the partition function  (\ref{16a}), (\ref{17})
with the obvious change $\beta^2\to x^2$. In this way we find that the random variable $\mu_N(x)$  is distributed as the product of two independent factors: $\mu_N(x)=e^{x\,u\sqrt{-2\ln{L}}}\tilde{\mu}_N(x)$, with the standard normal $u$ and $\tilde{\mu}_N(x)$ characterized by the
integer moments
\be \label{m11a}
\fl \mathbb{E}\left\{\tilde{\mu}_N(x)^p\right\}|_{\log{N_L}\gg 1}\approx \tilde{\mu}_e(x)^p \prod_{j=1}^{j=k} \frac{\Gamma^2[1-(j-1)
x^2] \Gamma(1-j
x^2)}{\Gamma[2-(k+j-2) x^2]}, \quad 0<x^2<1/p
\ee
Here the 'typical' value $\tilde{\mu_e}(x)$  is related to the similar scale for the full-circle case  (\ref{m12}):
$ \tilde{\mu}_e(x)=\frac{1}{(2\pi)^{x^2}}\mu_e(x)$, so shares the same multifractal scaling of the length of the intervals supporting high values.
To restore the probability density ${\cal P}(\xi)$ of the random variable $\xi=\tilde{\mu}_N(x)/\tilde{\mu}_e(x)$ we  define the generic moments $M_{x}(s)=\mathbb{E}\left\{\xi^{1-s}\right\}$ for any complex $s$ at fixed $0<x<1$. An explicit expression for $M_{x}(s)$ was found in
 \cite{FLeDR09} and \cite{Ost09}, and is given by (\ref{MT}). This allows us to represent ${\cal P}(\xi)$
 as a contour integral (\ref{FLD}). The dominating features of that distribution, like the powerlaw tail at $\xi\gg 1$,
 are the same as for the full-circle case.  Numerical verifications of these features are expected to be even more challenging, because by definition one is restricting attention to a small fraction of data that is itself hard to obtain in quantities sufficient for statistical analysis.

 \section{Motivation for the predictions relating to the extreme value statistics of $\zeta(1/2+it)$}
The calculations outlined in the previous section are based on estimating the asymptotics of the moments of the characteristic polynomials, and using the high moments to determine the extreme values.  The connections between these moments and those of the zeta function are now relatively well understood, at least conjecturally \cite{KS, CFKRS03, CFKRS05, GHK}.  For example, to leading order as $T\rightarrow\infty$
\be\label{KS}
\frac{1}{T}\int_0^T|\zeta(1/2+it)|^{2\lambda}dt\sim  a(\lambda)\mathbb{E}|p_N(\theta)|^{2\lambda},
\ee
with $N=\log\frac{T}{2\pi}$, where $a(\lambda)$ is defined by (\ref{arith}).
This allows us to use the previous calculations to motivate the predictions for the extreme value statistics for the zeta function.  The approach follows closely that already detailed in the random-matrix context, and so we  limit ourselves to outlining the principal steps.

The analogue of the partition function ${\cal Z}_{N}(\beta;L)$ defined by (\ref{1}) is clearly
\be\label{zeta1}
\frac{1}{2\pi}\log\frac{t}{2\pi}\int_t^{t+L}|\zeta(1/2+iy)|^{2\beta}dy
\ee
and so the analogue of the moments of the partition function (\ref{3}) is given by
\be
\fl
%\left(\frac{1}{2\pi}\log\frac{t}{2\pi}\right)^k
\frac{1}{T-T_0}\int_{T_0}^T\left(\frac{1}{2\pi}\log\frac{t}{2\pi}\right)^k\int_t^{t+L}\ldots \int_t^{t+L} \left\{ |\zeta(1/2+iy_1)|^{2\beta}\ldots |\zeta(1/2+iy_k)|^{2\beta}\right\}\prod_{j=1}^kdy_jdt
\ee
Interchanging the $t$-integral with the $y_j$-integrals (using the fact that the premultiplying $\log t$ factor is slowly varying) produces an integrand of the form
\be
\frac{1}{T-T_0}\int_{T_0}^T |\zeta(1/2+it+iy_1)|^{2\beta}\ldots |\zeta(1/2+it+iy_k)|^{2\beta}dt.
\ee
This is the analogue of the left-hand side of (\ref{4}).  A general expression for shifted moments of this kind was conjectured in \cite{CFKRS03, CFKRS05}.   This takes the form of a multiple integral; for example,
\begin{eqnarray}\label{eq:zetamomfull}
&&\frac{1}{T}\int_0^T|\zeta(1/2+it)|^{2k} dt =\int_{0}^T
\frac{(-1)^k}{k!^2}\frac{1}{(2\pi i)^{2k}}\\
&&\qquad\qquad\qquad\qquad \times  \oint\cdots \oint
\frac{G_{\zeta}(z_1, \dots,z_{2k})\Delta^2(z_1,\dots,z_{2k})} {
\prod_{j=1}^{2k}
z_j^{2k}}\nonumber\\
&&\qquad\qquad\qquad\qquad\times e^{\frac {1}{2}\log \frac{t}{2
\pi} \sum_{j=1}^{k}z_j-z_{k+j}}\,dz_1\dots dz_{2k} \,dt+
o(1),\nonumber
\end{eqnarray}
where
\begin{equation}
G_{\zeta}(z_1,\dots,z_{2k})= A_k(z_1,\dots,z_{2k})
\prod_{i=1}^k\prod_{j=1}^k\zeta(1+z_i-z_{k+j}) ,
\end{equation}
and $A_k$ is another Euler product which is analytic in the
regions in which we are interested.
The leading order asymptotics of such integrals was shown in \cite{CFKRS03, CFKRS05, Klost, Chand} to take a form analogous to (\ref{6}), but with an additional factor $a(\beta)^k$, where $a(\beta)$ is defined by (\ref{arith}).
The calculation then proceeds exactly as in the previous section; for example, the arithmetic factor multiples ${\cal Z}_e$.  In determining the extreme value statistics, however, the fact that the freezing transition occurs at $\beta=1$ and that $a(1)=1$, suggests that the arithmetical factor does not contribute at leading order.  In the computation of the measure of large values this factor remains.

It should be emphasized that this calculation only concerns the leading order asymptotics.  It is known that the lower order terms contribute significantly to the moments and are necessary to model numerical computations.  This is because the moments are long-range statistics and so are more sensitive to lower-order corrections than local statistics.  It would presumably be important to develop a method to calculate these lower order terms in order to develop a more accurate model for $p(x)$ at a finite height up the critical line.

 \section{Acknowledgements}
We are most grateful to Dr Ghaith Hiary and Mr Timoth\'{e}e Wintz for collaboration and extremely helpful discussions at various stages of this research, and for producing extensive numerical data supportive of our speculations.   YVF was supported by EPSRC grant EP/J002763/1 ``Insights into Disordered Landscapes via Random Matrix Theory and Statistical Mechanics''.  JPK was supported by a grant from the Leverhulme Trust and by the Air Force Office of Scientific Research, Air Force Material Command, USAF, under grant number FA8655-10-1-3088. The U.S. Government is authorized to reproduce and distribute reprints for Governmental purpose notwithstanding any copyright notation thereon.

\vspace{1cm}

\noindent  {\bf Appendix A:  Small-distance behaviour of the two-point correlation function of $\mbox{Re}\log{\zeta\left(\frac{1}{2}+it\right)}$.}\\

We have from the Euler product that
\be\label{zetaeuler}
\log{}\zeta(s)=-\sum_p\log\left(1-\frac{1}{p^s}\right)=\sum_p\sum_{n=1}^{\infty}\frac{1}{np^s}
\ee
where the $p-$summation includes all prime numbers. According to the Prime Number Theorem, the number $\pi(x)$ of primes smaller than $x$ grows as $\pi(x)\sim \frac{x}{\log{x}}$ when $x\to \infty$. This can be interpreted as  implying that the probability that a large integer $n$ is prime is asymptotically $1/\log{n}$, see \cite{H-B} for an introduction.  Moreover, sums over primes of the type $\sum_{p}f(p)$ that are dominated by large primes can be approximated by $\int^{\infty} \frac{f(x)}{\log{x}}\,dx$.

Substituting (\ref{zetaeuler}) into the two-point correlation function we obtain
\be\label{corzeta}
\fl \left\langle  V_t^{(\zeta)}(x_1)V_t^{(\zeta)}(x_2) \right\rangle= \sum_{p_1,p_2}\sum_{n_{1,2}=1}^{\infty}\frac{1}{n_1n_2}\frac{1}{p_1^{n_1/2}p_2^{n_2/2}}
\left\langle \cos{\left[n_1(t+x_1)\log{p_1}\right]}\cos{\left[n_2(t+x_2)\log{p_2}\right]}\right\rangle
\ee
Further rewriting the product of cosine factors as
\[
\fl \frac{1}{2}\cos{\left[t\log{\frac{p_1^{n_1}}{p_2^{n_2}}}+n_1\,x_1\,\log{p_1}-n_2\,x_2\,\log{p_2}\right]}+
\frac{1}{2}\cos{\left[t\log{p_1^{n_1}p_2^{n_2}}+n_1\,x_1\,\log{p_1}+n_2\,x_2\log{p_2}\right]}
\]
we see that second term will always give rise to rapid oscillations with $t$, whereas the first term does the same except when
$p_1^{n_1}=p_2^{n_2}$. We conclude that the dominating contributions come from those non-oscillating terms in the sum: $p_1=p_2=p; n_1=n_2=n$, resulting in the following ``diagonal approximation" for the two-point correlation function:
 \be\label{corzetadiag}
\left\langle  V_t^{(\zeta)}(x_1)V_t^{(\zeta)}(x_2) \right\rangle \approx  \frac{1}{2} \sum_{p}\sum_{n=1}^{\infty}\frac{1}{n^2p^n}\cos{\left[n\,(x_1-x_2)\,\log{p}\right]}
\ee
Using $\frac{1}{n^2}=\frac{1}{n}+\frac{1-n}{n^2}$ and denoting $x=|x_1-x_2|$ the relation(\ref{corzetadiag}) can be rewritten as
\be\label{corzetadiag1}
 \fl \left\langle  V_t^{(\zeta)}(x_1)V_t^{(\zeta)}(x_2) \right\rangle\approx  \frac{1}{2} \sum_{p}\sum_{n=1}^{\infty}\frac{1}{np^n}\cos{\left[n\,x\,\log{p}\right]}
 +\frac{1}{2} \sum_{p}\sum_{n=2}^{\infty}\frac{1-n}{n^2p^n}\cos{\left[n\,x\,\log{p}\right]}
\ee
Now notice that for any $n\ge 2$ the sums $\sum_{p}\frac{1}{p^n}$ are convergent, as the integrals $I_n=\int^{\infty} \frac{1}{x^n\log{x}}\,dx$
are convergent. This fact implies that the second sum in (\ref{corzetadiag1}) has a finite value when $x\to 0$. As to the first sum,
invoking (\ref{zetaeuler}) one observes that it can be related exactly to the Riemann zeta-function $\zeta{(s)}$ along the line $s=1+ix$, that is
$\frac{1}{2} \mbox{Re}\log{\zeta\left(1+ix\right)}$. Recall finally that $\zeta\left(1+ix\right)|_{x\to 0}\approx \frac{1}{ix}$,
implying in the limit $x\to 0$  the relation
\be\label{corzetadiag3}
  \left\langle  V_t^{(\zeta)}(x_1)V_t^{(\zeta)}(x_2) \right\rangle\approx  \frac{1}{2} \mbox{Re}\log{\zeta\left(1+i|x_1-x_2|\right)}+O(1)\approx -\frac{1}{2}\log{|x_1-x_2|}
\ee
Of course, the logarithmic form of correlations cannot hold for arbitrary small $x=|x_1-x_2|$; $\left\langle  \left[V_t^{(\zeta)}(x)\right]^2 \right\rangle$ must be obviously finite.
As is well known (see, e.g., \cite{Keat93})  the diagonal approximation works only for $p<\frac{t}{2\pi}$, and  breaks down for larger primes.
Taking this into account one can show that (\ref{corzetadiag3}) holds as long as $|x_1-x_2|\gg \frac{1}{\log{t}}$, whereas
 $\left\langle \left[V_t^{(\zeta)}(x)\right]^2 \right\rangle_t\approx \frac{1}{2}\log{\log{t}}$ for $t\to \infty$.

 The fact that $V_t^{(\zeta)}(x)$ has a gaussian distribution follows from a similar analysis of the higher moments.  Essentially, the diagonal terms dominate and so the prime sums representing the higher moments reduce to powers of the second moment in the same way as those of a gaussian.

\end{document}